\documentclass[useAMS,usenatbib]{mn2e}
\usepackage{epsfig}
\usepackage{graphicx}
\usepackage{amsmath}

\title[Factors influencing the emission-line response]{Interpreting broad emission-line variations I : Factors influencing the emission-line response}
\author[Goad and Korista]{M. R. Goad$^{1}$\thanks{E-mail: mg159@le.ac.uk}
and K. T. Korista$^{2}$.\\
$^{1}$Department of Physics and Astronomy, 
College of Science and Engineering, University of Leicester,  University Road, Leicester, LE1 7RH\\
$^{2}$Department of Physics, Western Michigan University,  Kalamazoo, Michigan 49008-5252, USA\\}

\begin{document}
\date{Received xxx; in original form
  June 2014}
\pagerange{\pageref{firstpage}--\pageref{lastpage}} \pubyear{2014}
\maketitle
\label{firstpage}
\begin{abstract}
We investigate the sensitivity of the measured broad emission-line
responsivity $d\log f_{\rm line}/d\log f_{\rm cont}$ to continuum
variations in the context of straw-man BLR geometries of varying size with
fixed BLR boundaries, and for which the intrinsic emission-line
responsivity is known a priori.  We find for a generic emission-line
that the measured responsivity $\eta_{\rm eff}$, delay and maximum of
the cross-correlation function are correlated for characteristic
continuum variability timescales $T_{\rm char}$ less than the maximum
delay for that line $\tau_{\rm max}$(line) for a particular choice of
BLR geometry and observer orientation. The above correlations are
manifestations of geometric dilution arising from reverberation
effects within the spatially extended BLR. When present, geometric
dilution reduces the measured responsivity, delay and maximum of the
cross-correlation function.  Conversely, geometric dilution is
minimised if $T_{\rm char} \ge \tau_{\rm max}$(line).  We also find
that the measured responsivity  and delay show a strong dependence on
light-curve duration, with shorter campaigns resulting in smaller than
expected values, and only a weak dependence on sampling rate (for
irregularly sampled data).

The observed strong negative correlation between continuum level and
line responsivity found in previous studies cannot be explained by
differences in the sampling pattern, light-curve duration or in terms
of purely geometrical effects. To explain this and to satisfy the
observed positive correlation between continuum luminosity and BLR
size in an individual source, the responsivity-weighted radius must
increase with increasing continuum luminosity. For a BLR with fixed
inner and outer boundaries this requires radial surface emissivity
distributions which deviate significantly from a simple power-law, and
in such a way that the intrinsic emission-line responsivity increases
toward larger BLR radii, in line with photoionisation calculations.

\end{abstract}
\section{Introduction}

Determining the geometry and kinematics of the broad emission-line
region (hereafter BLR) has been a long sought after goal of Active
Galactic Nuclei (AGN) monitoring campaigns. Early campaigns focused on
recovery of the 1-d response function $\Psi(\tau)$ for the broad
emission-lines, the function which maps the continuum variations onto
the line variations as a function of time-delay $\tau$, in an effort
at mapping the spatial distribution of the variable line emitting
gas. However, it was realised almost from the outset, that the form of
the 1-d response functions for diverse BLR geometries are largely
degenerate and that $\Psi(\tau)$ is unable to unambiguously pin down
the geometry of the BLR gas (Welsh and Horne 1991; P\'{e}rez, Robinson
and de la Fuente 1992b). Despite this, reverberation mapping has
enjoyed enormous success. Even with low quality data, measurement of
the delay between the continuum and broad emission-line variations via
the cross-correlation function (hereafter CCF), yields with a few
assumptions, a measure of the luminosity-weighted size of the
line-emitting region. When combined with a measure of the velocity
dispersion of the line-emitting gas, and assuming that the BLR gas is
virialised, broad line variability data can be used to determine the
mass of the central black hole.  Indeed RM mass estimates are now
available for more than 40 nearby AGN, from which scaling relations
have been derived for several UV and optical broad emission-lines
allowing access to black hole mass determinations for AGN from single
epoch spectra (Kelly and Bechtold 2007; Denney et al. 2009; Shen et
al. 2008; Shen and Kelly 2010; Runnoe et al. 2013).

Applying time variability studies to multiple lines in individual
sources has proven equally profitable. For the best studied source
NGC~5548, reverberation mapping experiments reveal the BLR to be both
spatially extended and highly stratified, with a broad range in delays
exhibited among lines of differing ionisation stage suggestive of
strong gradients in density and/or ionisation (e.g. Netzer and Maoz
1990; Krolik et al. 1991; Clavel et al. 1991; Peterson et al. 1992,
1994, 2002, Korista et al. 1995, and references therein).

 The shortest
delays ($\sim$ a few days) are measured for the UV high ionisation
lines (HILs), (e.g. N~{\sc v}, C~{\sc iv}, He~{\sc ii}). For these
lines the recovered 1-d response functions are temporally unresolved
(owing to sparse sampling of the UV light-curves), peaking at
zero-delay, and declining rapidly on timescales of a few days. This
contrasts with the recovered response functions for the LILs. For
example, the broad optical recombination lines (H$\alpha$, H$\beta$),
have little response at zero delay, instead rising to a peak at
$\sim$20 days before declining rapidly toward zero (Horne, Welsh and Peterson 1991; Ferland et al. 1992). While the UV Fe~{\sc
  ii} emission-lines were found to vary on a similar timescale and with
similar amplitude to the Balmer emission-lines (Maoz et al. 1993), for
Mg~{\sc ii} and the optical Fe~{\sc ii} lines, the variability
amplitude was so small that only a lower limit on the delay was
possible (Clavel et al. 1991; Vestergaard and Peterson
2005). Similarly, for C~{\sc iii]}, a density sensitive
  inter-combination line, only a lower limit for the delay is
  available, suggesting that the characteristic density of the BLR
  decreases radially outward (Clavel et al. 1991; Krolik et al. 1991).

\subsection{The 13 yr ground-based optical monitoring campaign of NGC~5548}

Of the $\approx$40 or so nearby AGN with extensive ground-based
continuum--emission-line monitoring data, the 13 yr ground-based
monitoring campaign on NGC~5548 carried out by the Ohio State University as part
of the AGN Watch collaboration remains the de-facto gold standard,
both in terms of sampling frequency, campaign duration, and data
quality (Peterson et al. 2002 and references therein). While not
approaching the signal to noise or sampling frequency available with
space-based instruments, the extensive ground-based coverage allows
investigation of the emission-line delay and amplitude of response on
a season by season basis, and on timescales longer than the BLR
dynamical timescale ($\sim$ a few years for NGC~5548), and
importantly, their dependence on ionising continuum luminosity.

\begin{figure}
\resizebox{\hsize}{!}{\includegraphics[angle=0,width=8cm]{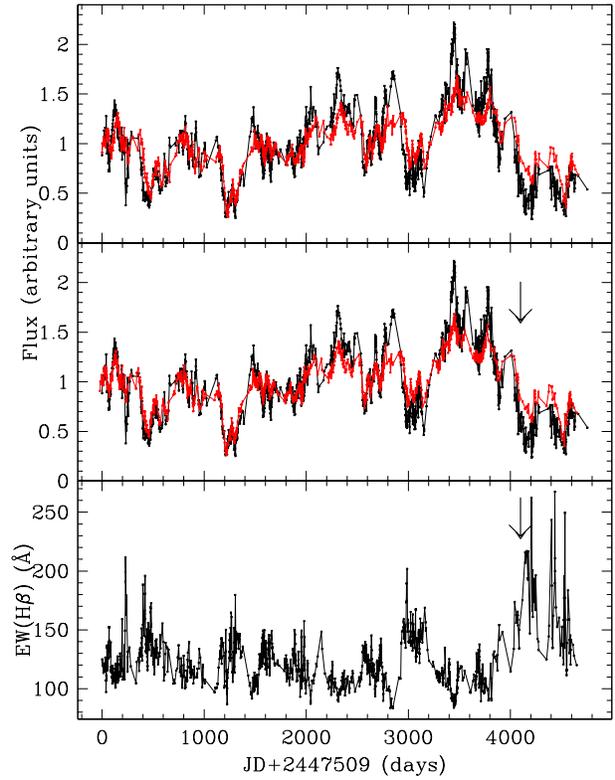}}
\caption{Top panel - the galaxy subtracted optical continuum and 
narrow line subtracted broad H$\beta$ emission-line light-curve of
NGC~5548 as observed during the 13~yr optical monitoring campaign. For
clarity both light-curves have first been normalised to their
respective mean values (see text for details).  Middle panel, as
above, but with each season of data corrected for the measured delay
between the continuum and emission-line variations. Lower panel - the
time-variable EW for broad H$\beta$ as determined from the lag
corrected data. The downward arrow indicates the onset of large EW
which follows a period of prolonged high continuum and emission-line
fluxes (\S1.1.)}\label{fig1}
\end{figure}

In Figure~\ref{fig1} we show the galaxy subtracted optical (5100\AA)
continuum light curve (black points) and the corresponding broad
H$\beta$ emission-line light curve (red points) for NGC~5548 after
having first corrected for contaminating narrow H$\beta$ emission
using the latest values of the (now known to be time-variable on
timescales of $<$ 13 years) narrow line contribution to broad H$\beta$
from Peterson et al. (2013). For the purposes of illustration only we
have normalised the continuum and emission-line light-curves
to their respective mean values.  The broad
emission-line light-curve clearly follows the continuum light-curve
albeit with a small delay (see Peterson et al. 2002, and references
therein). In the middle panel we show the galaxy subtracted optical
(5100\AA) continuum light curve (black points) together with the
narrow line subtracted broad H$\beta$ emission-line light-curve, with
the latter shifted, on a season by season basis, by the delay (as
measured from the centroid of the cross-correlation function,
hereafter CCF) between the continuum and broad emission-line light
curve, using values taken from Peterson et al. 2002). The continuum
and emission-line light curves are well-matched, with the optical
continuum light curve displaying larger amplitude variations than the
broad emission-line light-curve.  In the lower panel of
Figure~\ref{fig1} we show the time variable EW of the broad H$\beta$
emission line light-curve. Since the continuum and broad emission-line
light-curves are irregularly sampled and the broad emission-line
light-curve has been shifted in time, we determine the line EW (the
ratio of the line to continuum flux) at the corresponding continuum
values by linearly interpolating between nearest neighbour points.

Figure~\ref{fig1} indicates several key observational
features of AGN monitoring campaigns. First, the broad emission-line
light-curve correlates well with the optical continuum light curve
which not only confirms ionising continuum variations as the key
driving mechanism for the broad emission-line variations, but suggests
that the {\em continuum variability timescale and BLR size are generally
well-matched\/}, and furthermore indicates that the optical continuum may
be used as a proxy for the driving UV ionising continuum\footnote{If
the UV and optical continuum arise from the purported accretion disc,
then the expectation is that the optical continuum originates at larger
disc radii than the UV continuum, and will consequently display more
slowly varying smaller amplitude variations.}.

However, when looked at in detail, the broad H$\beta$ emission-line
light-curve shows small seasonal shifts (in delay) of varying size
relative to the optical continuum light curve, and is somewhat
smoother in appearance, displaying smaller amplitude excursions about
its mean level. Both of these effects are normally attributed to
light-travel time-effects (reverberation) within the spatially
extended BLR. Both the continuum--emission-line delay (or lag) and the
amplitude of the emission-line variations (the line responsivity,
$\eta$) are key components of the broad emission line response
function $\Psi(\tau)$, the convolution kernel relating the continuum
to broad emission-line variations the recovery of which has been the
major goal of AGN monitoring (reverberation mapping, RM) campaigns
over the past 25 years.

However, there is a subtlety revealed by the continuum and
emission-line light-curves which is often overlooked. When recast in
terms of the line EW, it becomes apparent that the line EW for broad
H$\beta$ is not constant, but instead varies by a factor of $\approx
2$ over the full 13~yr campaign. Furthermore, the H$\beta$ EW varies
inversely with the continuum level, with the largest values occurring
during low continuum states and the smallest values occurring during high
continuum states.
Formally, the measured continuum -- broad
emission-line fluxes in a given AGN are related by

\begin{equation}
\label{blah}
f(H\beta) \propto f_{cont}^{\eta_{\rm eff} } \, , 
\end{equation}

\noindent where $\eta_{\rm eff}$, the power-law index in this
relation, measures the effective responsivity of a particular
emission-line over the full BLR, modulo a first order correction for
the continuum--emission-line delay, which dilutes the signal
(geometric dilution) and introduces scatter in the above
relation\footnote{If the BLR is spatially extended, the transfer
  function may have an extended tail, such that the longest delays are
  significantly larger than the typical continuum variability
  timescale. Light-travel time effects (reverberation) across the
  spatially extended BLR will then act to reduce the amplitude of the
  emission-line response. We refer to this effect as geometric
  dilution. If the tail of the response function is sufficiently
  large, then simply correcting for the emission-line lag may not be
  enough to remove the effect of geometric dilution on the
  measured line responsivity.}. In terms of the line EW,
equation~\ref{blah} becomes

\begin{equation}
EW(H\beta) = \frac{f(H\beta)} {f_{cont}}  \propto f_{cont}^{\eta_{\rm eff}-1} \, .
\end{equation}

\noindent 
Since in general we measure a local equivalent width for H$\beta$
using nearby continuum bands (typically 5100\AA), this simplified
expression ignores the fact that the driving UV continuum may exhibit
larger amplitude variations than the optical continuum bands.  Values
of $\eta_{\rm eff} < 1$ may be associated with an intrinsic Baldwin
effect (e.g. Kinney et al. 1990; Pogge and Peterson 1992; Gilbert and
Peterson 2003; Han et al. 2011) for this line, and the EW(H$\beta$) is
indeed found to be inversely correlated with continuum strength, being
larger in low continuum states (compare the middle and lower panels of
Figure~\ref{fig1}), and smaller in high continuum states (Gilbert and
Peterson 2003; Goad, Korista and Knigge 2004; Korista and Goad 2004).
Since the emission-line EW is a measure of the efficiency with which
the BLR gas reprocesses the incident ionising continuum into line
emission, this suggests that the continuum re-processing efficiency
for H$\beta$ decreases as the ionising continuum strength increases
(Korista and Goad 2004). Indeed the measured factor of 2 or more
variation in emission-line EW for broad H$\beta$ relative to the
optical continuum band in NGC~5548 is in close agreement with the
predicted variation in EW(H$\beta$) from photoionisation model
calculations (see \S3.2 of Korista and Goad 2004). Note that while the
continuum re-processing efficiency for a single line may change
dramatically with continuum level (for example if $\eta_{\rm eff}< 1$,
we will observe an intrinsic Baldwin effect), the overall gas
re-processing efficiency may remain unchanged if contributions from
other lines adjust accordingly (Maoz 1992)\footnote{If for example the
  ionising continuum shape changes, but the number of hydrogen
  ionising photons remains the same, certain lines may show enhanced
  re-processing efficiency, due to a change in the ionisation state of
  the gas, at the expense of others, while the integrated response
  (summing over all lines) remains unchanged. This scenario has
  previously been used to explain the unusually strong response of the
  C~{\sc iv} emission-line in the latter stages of the 1989 IUE
  monitoring campaign of NGC~5548.}.

Gilbert and Peterson (2003), and Goad, Korista and Knigge (2004) went
on to show that the measured line responsivity $\eta_{\rm eff}$ is not
constant, but instead varies with continuum state, being smaller in
high continuum states, and larger in low continuum states, a
continuum-level dependent emission-line responsivity $\eta_{\rm eff} =
\eta_{\rm eff}(L_{\rm cont}(t))$. Since the gas covering fraction is
unlikely to correlate strongly with ionising continuum strength, then
these observations point towards a physical origin for this effect
within the BLR gas. Notice that the largest H$\beta$ EWs in the lower
panel of Figure~\ref{fig1} correspond to low continuum states and
indicate a higher re-processing efficiency for the line-emitting gas at
lower ionising continuum fluxes, as predicted by the photoionisation
calculations of Korista and Goad (2004) and equation~2, for $\eta_{\rm
  eff} < 1$.

The large upward and downward excursions in line EW for broad H$\beta$
seen in the lower panel of Figure~1 anti-correlate with continuum
level (Goad, Korista and Knigge 2004) and can be explained in terms of
(i) a decrease in the continuum re-processing efficiency for H$\beta$
with increased continuum flux as predicted by photoionisation model
calculations, a purely local effect (e.g. Korista and Goad 2004, and
see \S2.1), (ii) the increase in the luminosity-weighted radius with
increasing continuum flux which subsequently follows, and which for a
given characteristic continuum variability timescale $T_{\rm char}$
results in a larger delay and a lower amplitude emission-line response
(here referred to as geometric dilution), and (iii) hysteresis effects
arising from the finite light-crossing time of a spatially extended
BLR.  The larger H$\beta$ EW found for low continuum states, and in
particular in those that follow prior high continuum states (for
example, as indicated by the arrow in the middle and lower panels of
Figure~1) may in part be attributed to hysteresis effects. While an
external observer sees the continuum decline with no delay, the line
emission arises from gas distributed over a broad range in delay and
will be largely dominated by contributions from gas at larger radii
responding to the prior (higher) continuum states. Hysteresis in the
time-variable line EW may therefore be used as indicator of a
spatially extended BLR.

The decline in the continuum re-processing efficiency for a given line
with increasing continuum flux, and the resulting increase in the
lines' luminosity-weighted radius with continuum level, are associated with
what is commonly referred to as ``breathing'' (e.g. Goad et al. 1993;
Korista and Goad 2004; Cackett and Horne 2006).
We explore this effect in a forthcoming paper.  Cackett
and Horne (2006) provided strong supporting evidence for a breathing
BLR in NGC~5548, showing that a time-variable luminosity-dependent
response function $\Psi(\tau,L(t))$ provides a better fit to the 13~yr
broad H$\beta$ emission-line light curve for this source.

Finally, it is interesting to note that there is an apparent lower
limit to the measured time-variable EW for broad H$\beta$ of $\approx$
80--100~\AA\ (lower panel of Figure~\ref{fig1}). Additionally, the
time-variable line EW for broad H$\beta$ does not appear invariant to
rotation through 180 degrees\. That is, if we invert the H$\beta$ EW
light-curve it does not have the same functional form. This may in
part be explained by the general finding that the emission-line
responsivity is larger in low continuum states than in high continuum
states (e.g. Korista and Goad 2004) and which introduces asymmetry
into the line EW variations with continuum level. The apparent bounded
behaviour of broad H$\beta$ line EW in high continuum states (and the
possibility of a similar bounded behaviour in low continuum states)
will be explored further in paper~{\sc ii}.

\subsection{The power of photoionisation modelling}

Though RM campaigns have been hugely successful, an unambiguous
interpretation of the recovered response function remains difficult,
because in general, {\em there is no simple one-one relationship
between BLR size $R$ and the measured time-delay $\tau$\/} and because
the importance of the local gas physics in determining the amplitude
of the emission-line response to continuum variations (the line
responsivity) has with a few exceptions been largely
overlooked. $\Psi(\tau)$ far from being a simple entity, is determined
among other things by the BLR geometry (e.g. sphere, disc, bowl etc),
the amplitude and characteristic timescale of the driving continuum
light-curve, and by properties of the local gas physics, many of which
may themselves vary in a time-dependent manner (for example, as a
function of incident continuum flux), all of which are then moderated by our
ability to sample the continuum light-curves with sufficient frequency
and over long enough duration to mitigate against windowing and
sampling effects\footnote{While the velocity resolved response
function $\Psi$($v$,$\tau$) breaks the degeneracy inherent among
different BLR geometries (e.g. Welsh and Horne 1991; P\'{e}rez,
Robinson and de la Fuente 1992), similar arguments with regards its
recovery and interpretation also apply.}.

As an exemplar, when taken at face-value the short response timescales
found for the HILs in NGC~5548 suggests that these lines either form
at small BLR radii, and/or originate close to the line of sight to the
observer.  However, as their name implies, the HILs require a more
intense radiation field for a given gas hydrogen density than the
LILs, and therefore the most plausible explanation is that these lines
form at small BLR radii, where the radiation field is more
intense\footnote{A possible exception to this simple rule would arise if
  the characteristic hydrogen gas density $n_{\rm H}$ falls faster than
  $1/r^2$.}. Thus an understanding of the photoionisation physics of
the gas allows us to distinguish these two scenarios.  In a similar
fashion, the absence of response in the optical recombination on short
timescales, when interpreted geometrically, suggests an absence of gas
along the observers line of sight, and by implication a departure from
spherical symmetry.  However, photoionisation models indicate that the
gas emitting the optical hydrogen recombination lines may be optically
thick in the line (Ferland et al 1992; O'Brien, Goad and Gondhalekar
1994; Korista and Goad 2004). If so, the absence of response on short
timescales, far from indicating an absence of gas, rather suggests
that these lines preferentially emerge in the direction of the
illuminating source, and therefore the gas remains unseen (for these lines).

The importance of using photoionisation models in the interpretation
of broad emission-line variability data is clear (see Goad et al 1993,
O'Brien et al 1994,1995; Korista and Goad 2004; Horne, Korista and
Goad 2003).  In this contribution we use a forward modelling approach
to investigate, in a controlled manner, the relative importance of
several key factors underpinning the measured emission-line response
amplitude (responsivity) and delay (or lag) in response to ionising
continuum variations.  We begin in \S2 by introducing a framework for
discussing the emission-line responsivity, distinguishing between the
in-situ gas responsivity derived from photoionisation calculations and
the effective line responsivity, $\eta_{\rm eff}$, measured by the
observers. We then implement an alternative means of measuring
$\eta_{\rm eff}$ which doesn't require a correction for the
continuum--emission-line delays (Krolik et al. 1991).

In \S\ref{sims} we introduce a set of controlled reverberation mapping
experiments in which we investigate the sensitivity of the {\em
  measured} emission-line response amplitude (the emission-line
responsivity $\eta_{\rm eff}$) and emission-line delay $\tau$ to the
local gas physics, and BLR geometry (radial extent and geometric
configuration), for a range of BLR geometries centred about the BLR
size and continuum luminosity normalisation for the well-studied AGN
NGC 5548, driven by continuum light-curves with differing
characteristic timescales and variability amplitudes, and moderated
by the campaign duration and sampling frequency.

In \S\ref{discuss} we place our work in context of the observed
variability behaviour of the broad H$\beta$ emission-line in NGC~5548
addressing in particular the implications of adopting a spatially and
temporally dependent emission-line responsivity (ie. ``breathing'')
and outline several avenues for further investigation where
significant future progress can be made.

\section{The emission-line responsivity : definitions}
\subsection{The in-situ gas responsivity}

From a photoionisation modelling perspective we are primarily
interested in the in-situ microscopic physics of the line emitting gas
and variations in the locally emergent emission-line intensities $F_{\rm line}$,
  about their equilibrium values resulting from small changes in the
  incident hydrogen ionising photon flux $\Phi_{\rm H}$, normally
  referred to as the emission-line responsivity $\eta$. Formally,
  $\eta$ can be written

\begin{equation}
\eta=\frac{d\log F_{\rm line}}{d\log \Phi_{\rm H}} \, .
\end{equation}

\noindent and is a measure (locally) of the re-processing efficiency 
of the line-emitting gas. Assuming that the spectral energy
distribution (SED) of the ionising continuum remains constant (to
first order), then this may be rewritten as

\begin{equation}
\eta=\frac{d\log EW}{ d\log \Phi_{\rm H}}+1  \, ,
\end{equation}

\noindent and can be computed directly from photoionisation model
grids of line EW as a function of hydrogen gas density $n_{\rm H}$ and
hydrogen ionising photon flux $\Phi_{\rm H}$ (e.g. Korista and Goad
2004).  This definition of line responsivity $\eta$ is useful because
it makes the minimum number of model-dependent assumptions, depending
only on the local gas physics (e.g. $\Phi_{\rm H}$, $n_{\rm H}$), and
importantly is independent of any assumed geometry, or indeed
weighting function describing the run of physical properties with
radius.  In previous work Korista and Goad (2004) showed that the
continuum re-processing efficiencies for the strong optical
recombination lines display a general inverse correlation with
incident hydrogen ionising photon flux, and consequently their line
responsivities tend to increase toward larger BLR radii.  Thus for
these emission-lines their measured responsivities $\eta_{\rm eff}$
may provide an additional constraint upon the run of gas physical
conditions with radius.

\begin{figure}
\resizebox{\hsize}{!}{\includegraphics[angle=0,width=8cm]{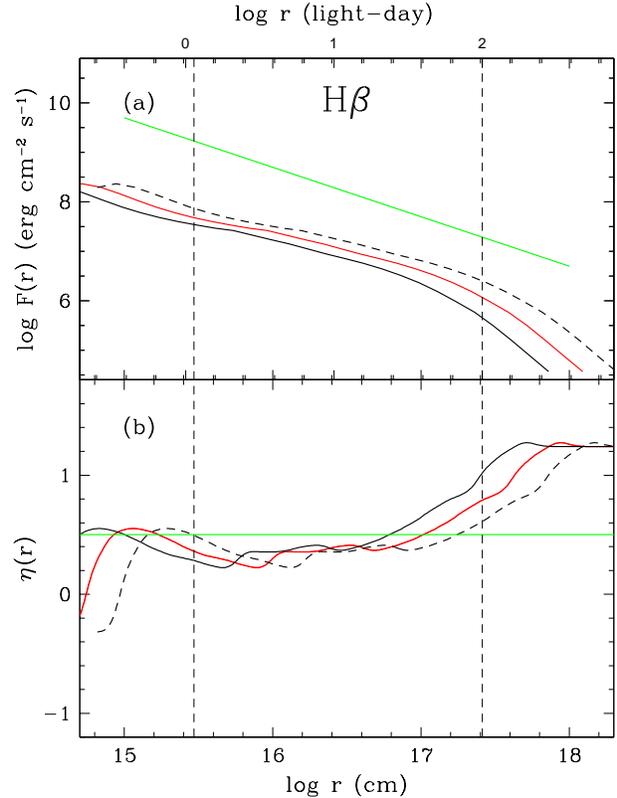}}
\caption{Upper panel - photoionisation model calculations of the
  radial surface emissivity distribution (solid red line) for broad
  H$\beta$ for an LOC model and continuum normalisation (see Korista
  and Goad 2004) appropriate for NGC~5548.
  Also shown are model low- (solid black line) and high-state (dashed
  black line) radial surface emissivity distributions for H$\beta$
  chosen to represent the peak-to-peak UV continuum variation in this
  source (factor 8.2 in flux).  The green line indicates a radial
  surface emissivity distribution with power-law index $\gamma=-1$,
  corresponding to a radial responsivity distribution
  $\eta(r)=-(\gamma/2)=0.5=constant\; \forall r$ and is used here
  simply to guide the eye. Lower panel - the equivalent radial
  responsivity distributions corresponding to changes in their
  emission-line fluxes about their equilibrium states in response to
  small continuum variations. The dashed vertical lines indicate the
  radial extent of our fiducial BLR model.}
\label{eta_hb}
\end{figure}


A radial surface emissivity distribution of an individual
emission-line $F_{\rm line}(r)$ (hereafter, $F(r)$) can be determined
by summing over a gas density distribution as a function of radial
distance $r$ (e.g., see Korista and Goad 2000)\footnote{In the Local
  Optimally-emitting Cloud (LOC) model of the BLR (Baldwin et
  al.\ 1995), $F(r)$ can be constructed by adopting a weighting
  function for the gas density distribution, often assumed to be a
  simple power-law, $g(n_{H}) \propto n_{H}^{-1}$. Krause et
  al. (2012) used magnetohydrodynamic simulations to show that in the
  presence of a helical magnetic field, fragmentation of broad-line
  clouds due to Kelvin-Helmholtz instabilities produces a power-law
  distribution in gas density, similar to that quoted above and
  applied in the LOC model calculations of Baldwin et al. (1995), and
  Korista and Goad (2000, 2001, 2004).}.  From $F(r)$ we can
subsequently calculate a radially dependent line responsivity
$\eta(r)$ (Goad, O'Brien and Gondhalekar 1993; Korista and Goad
2004). Formally, $\eta(r)$ can be written as


\begin{equation}
\label{eqtn_eta}
\eta(r)=\frac{d\log F(r)}{d\log \Phi_{\rm H}} \, ,
\end{equation}

\noindent and indicates the instantaneous variation in the radial
surface emissivity distribution $F(r)$ about its equilibrium value, to
small variations in the ionising continuum flux.

Since $\Phi_{\rm H}~\propto~r^{-2}$, $\eta(r)$ is given by $-0.5d\log
F(r)/d\log r$.  Example radial responsivity curves $\eta(r)$ for the
strong UV and optical broad emission-lines can be found in Goad et
al. (1993), and Korista and Goad (2004).  For illustration, we show in
Figure~\ref{eta_hb} (upper panel, red line) the radial surface line
emissivity for broad H$\beta$ (solid red line) determined for an LOC
model of NGC~5548 (Korista and Goad 2004). For broad H$\beta$, $F(r)$
deviates significantly from a single power-law, being better
represented by a broken power-law with power-law index $\gamma \approx
-0.7$ for radii less than $\approx$25~light-days and steeper than $-2$
for radii greater than $\approx$160~light-days\footnote{The radial
  surface emissivity curves and corresponding radial responsivity curves
  shown here assumes that gas at large radii remains grain free.},
corresponding to radial responsivities $\eta(r)$, of $\approx 0.35$
and $> 1$ respectively (see Figure~\ref{eta_hb} lower panel, solid red
line). For power-law radial surface emissivity distributions
$F(r)\propto r^{\gamma}$, $\eta(r) = -(\gamma/2)=constant$ $\forall$\/
$r$, ie. $\eta(r)$ is a constant both spatially and
temporally. However, since in the above example $F(r)$ is not a simple
power-law, $\eta(r)$ depends upon the amplitude of the continuum
variations, and is therefore luminosity dependent ie. $\eta(r, L_{\rm
  cont}(t))$.

We can also compute radial surface emissivity distributions $F(r)$ and
radial responsivity distributions $\eta(r)$ corresponding to small
continuum variations about hypothetical low- (solid black line) and
high- (dashed black line) equilibrium continuum states. The radial
line responsivity distributions resulting from larger continuum
variations, for example when traversing from a low- to high- continuum
state, will lie somewhere between the radial responsivity
distributions determined for small continuum variations about the low-
and high-continuum equilibrium states respectively (see e.g. Korista and Goad
(2004) for details).

\subsection{Measuring the emission-line responsivity, an observers' perspective}

As mentioned in \S1.1 (equation~1), the emission-line responsivity
$\eta_{\rm eff}$ is normally calculated from a measurement of the
slope in the relation $d\log f_{\rm line}/d\log f_{\rm cont}$, after
first applying a gross correction for the average delay $<\tau>$
between the continuum and emission-line variations (e.g. Krolik
et~al. 1991; Pogge and Peterson 1992; Gilbert and Peterson 2003; Goad,
Korista, and Knigge 2004), and having first adequately accounted for
contributions from non-varying components, for example, the host
galaxy contribution to the continuum light-curve and the contribution
of (non-)variable narrow emission-lines to measurements of the broad
emission-line flux. Clearly, when determining $\eta_{\rm eff}$, it is
important to ensure that the emission-line flux is referenced to the
correct (in time) continuum value. Pogge and Peterson (1992) suggested
that a global correction for the emission-line delay, $\tau$, is
sufficient to enable an accurate recovery of $\eta_{\rm eff}$,
demonstrating that data corrected in this fashion shows significantly
reduced scatter. However, we note that such a correction introduces
its own problems.  Krolik et~al. (1991) and Pogge and Peterson (1992),
when calculating $\eta_{\rm eff}$ for the broad UV emission-lines (as
observed with IUE during the 1989 AGN monitoring campaign of the
nearby Seyfert 1 galaxy NGC 5548), looked at regularly sampled data,
and avoided the need to interpolate their data by shifting their broad
emission-line light-curves by lags which were integer multiples of the
4-day sampling rate. By contrast, the unavoidable irregular sampling
of the 13~yr ground-based optical light-curves of NGC~5548 required a
different approach. Goad, Korista and Knigge (2004) applied the mean
delay ($<\tau>$) , as calculated from the centroid of the CCF, for
each of the 13 observing seasons of NGC~5548, to the H$\beta$
emission-line light-curve, reconstructing the corresponding optical
continuum flux from a weighted average of the continuum points
bracketing the observation, using weights and errors derived from the
1st order structure function of the continuum light-curve.  They found
that for the broad H$\beta$ emission-line, the effective line
responsivity $\eta_{\rm eff}$ {\em referenced to the optical
  continuum\/}, varies between 0.4 -- 1.0 over the 13 year optical
campaign, with $\eta_{\rm eff}$ inversely correlated with continuum
flux, such that the responsivity is generally larger at low continuum
flux levels (Goad, Korista and Knigge 2004, their
Figure~6)\footnote{Gilbert and Peterson (2003) showed that the
  measured range in $\eta_{\rm eff}$ ($\approx$0.53--0.65) for the
  full 13~yr ground-based optical monitoring campaign on NGC~5548
  depends upon the fitting process employed.}. Goad, Korista and
Knigge (2004, their Figure 4) suggested that the observed inverse
correlation between the broad emission-line responsivity and continuum
flux could best be explained in terms of photoionisation models. They
predict a larger emission-line responsivity at low incident continuum
fluxes, together with a more coherent response of the lines to
continuum variations of a given characteristic timescale $T_{\rm
  char}$ resulting from a smaller responsivity-weighted radius for a
BLR of fixed size.  A similar inverse correlation between the broad
emission-line responsivity and continuum flux has subsequently been
found for the broad UV emission-lines in the nearby Seyfert 1 galaxy
NGC~4151 (Kong et al. 2006), albeit over a much larger range in
continuum flux ($\sim$ factor of 100). This effect which is briefly
discussed in \S4 will be explored more fully in paper~{\sc ii}.

In previous work, Korista and Goad (2004) investigated how small
variations in the ionising continuum about its equilibrium value
induces local changes in the emission-line re-processing efficiency
leading to a radially dependent emission-line responsivity, $\eta(r)$.
Aside from this local effect, there are additional (non-local) effects
which act to modify the {\em intrinsic\/} responsivity and give rise
to {\em measured\/} values of the responsivity $\eta_{\rm eff}$ which
are generally smaller.  These non-local effects can be broadly
separated into properties of the system which are beyond the control
of the observer, and those which relate to how the system is measured
and over which the observer has some influence.  The former includes
for example, the characteristic timescale $T_{\rm char}$ and amplitude
$\sigma$ of the variable driving continuum light-curve, that together
with a given BLR geometry, size, and observer orientation conspire to
dilute the measured responsivity (geometric dilution). The latter
includes the duration $T_{\rm dur}$ and sampling rate $\Delta t$ of a
particular observing campaign. We show here that $T_{\rm dur}$ should
be carefully chosen in order to minimise the effect of windowing on
the measured emission-line responsivity.  We explore all of these
effects for several of the more familiar BLR geometries (e.g.
spherical, disc and bowl-shaped BLR geometries).

\subsubsection{$F_{\rm var}$, an alternative estimate of $\eta_{\rm eff}$.}\label{fvar}

Krolik et al. (1991) suggested an alternative and far simpler means of
estimating the effective line responsivity, using the ratio of the
fractional variability in the line and continuum $F_{\rm
  var}$(line)$/F_{\rm var}$(cont). For well-sampled long duration
light curves (so that all frequencies are suitably covered) $F_{\rm
  var}$(line)$/F_{\rm var}$(cont) is an unbiased estimator of the
slope of the response $\eta_{\rm eff}$ (their \S5.2), provided that
non-varying components have been adequately accounted for (e.g. the
host galaxy contribution to the continuum light-curve and the narrow
emission-line contribution to the broad emission-line light curve).

Formally, the fractional variability of a time-series,
$F_{\rm var}$  is given as

\begin{equation}
F_{\rm var} = \frac{ (\sigma^{2} - \Delta^{2})^{1/2} } {<f>} \, ,
\end{equation}

\noindent where the variance, $\sigma^{2}$, is given by

\begin{equation}
\sigma^{2} = \frac{1}{N-1}\sum_{i=1}^{N} (f_{i} - <f>)^{2} \, ,
\end{equation}

\noindent and $\Delta^{2}$, the mean squared error, is 
\begin{equation}
\Delta^{2} = \frac{1}{N}\sum_{i=1}^{N}\Delta_{i}^{2} \, 
\end{equation} 

\noindent where $\Delta_{i}$ is the uncertainty on the individual
measurements, and the mean, $<f>$, is written as

\begin{equation}
<f> = \frac{1}{N}\sum_{i=1}^{N} f_{i} \, .
\end{equation}

\begin{figure}
\resizebox{\hsize}{!}{\includegraphics[angle=0,width=8cm]{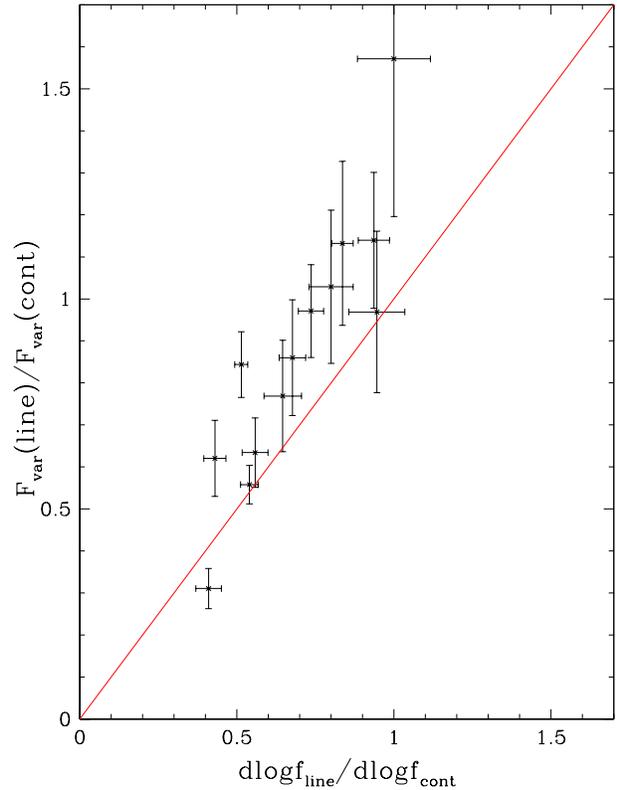}}
\caption{A comparison of the measured broad H$\beta$ emission-line
  response, for each of the 13 seasons of optical data for
  NGC~5548. $d\log f_{\rm line} /d\log f_{\rm cont}$ values are taken from Goad,
  Korista and Knigge (2004) with errors in the slope determined using
  the bootstrap technique, utilising 10000 simulations, with full
  replacement.  }
\label{brad}
\end{figure}

\noindent One advantage of using $F_{\rm var}$ is that it is
relatively straightforward to measure.  However its robustness as an
unbiased estimator of $\eta_{\rm eff}$ is unsubstantiated, and in
particular its sensitivity to geometric dilution and windowing effects
remain untested.  We discuss this further in \S\ref{sims}. In
Figure~\ref{brad} we show a comparison between $\eta_{\rm eff}$
derived from the ratio $F_{\rm var}(\rm line)/F_{\rm var}(\rm cont)$
and that found using the traditional method, $d\log f_{\rm line}/d\log
f_{\rm cont}$, both referenced to the optical continuum for each season
of the 13 yr NGC~5548 optical monitoring campaign. In each case, the
host galaxy contribution to the continuum light-curve and the narrow
emission-line contribution to the broad emission-line light curve have
been removed, while for the latter, we have also subtracted from each
season the mean delay, $<\tau>$, between the continuum and
emission-line variations (see Goad, Korista and Knigge 2004 for
details). While the two estimates of $\eta_{\rm eff}$ are not
identical, they do show a strong correlation, with estimates of
$\eta_{\rm eff}$ from the ratio of the fractional variation in the
line to the fractional variation in the continuum showing a larger
spread.

One obvious drawback to using $F_{\rm var}$ in the determination of
$\eta_{\rm eff}$ is that any information about the continuum
luminosity is subsequently lost.  However, this information may be
recovered simply by measuring $F_{\rm var}$ values over light-curve
segments corresponding to similar continuum states.

\section{Simulations}\label{sims}

When considering the measurement of parameters related to BLR
variability we must first separate out those effects which are related
to the properties of the driving continuum, e.g. amplitude and
characteristic variability timescale, from those governed by the local
gas physics, the BLR geometry and observer orientation, and the
observing window (e.g duration and sampling rate of the continuum and
emission-line light curves). The dependence of the
continuum--emission-line delay (or lag) on light-curve duration and
sampling window for BLRs of varying sizes has been studied elsewhere
(e.g. P\'{e}rez, Robinson and de la Fuente 1992a; Welsh 1999). To
summarise, these studies suggest that the continuum--emission-line
delay (or lag) is biased toward small BLR radii, a consequence of
finite duration of the observing campaigns and the presence of low
frequency power in the light-curves. This effect is more pronounced
for the peak of the CCF (the lag) than it is for the CCF centroid.
The cross-correlation function is simply the convolution of the
continuum autocorrelation function (a symmetric function) with the
response function.  Thus for light-curves of sufficient duration, the
centroid of the CCF (or luminosity-weighted radius) and the centroid
of the response function (the responsivity-weighted radius) are
equivalent (Penston 1991; Koratkar and Gaskell 1991; Goad, O'Brien and
Gondhalekar 1993).
Here we study the relationship between the
measurement of the emission-line responsivity $\eta_{\rm eff}$ and lag
for 3 of the more popular BLR geometries e.g. a sphere, a disc and
what we here refer to as our fiducial BLR geometry, a bowl-shaped
geometry (see \S3.1), within the context of prescribed differences in
the characteristic behaviour of the driving continuum light-curve.

\subsection{A fiducial BLR model}\label{fiducial}

In previous work, Goad, Korista and Ruff (2012), introduced a new
model for the BLR, one in which the BLR gas occupies a region bridging
the outer accretion disc with the inner edge of the dusty torus, the
surface of which approximates the shape of a bowl. This geometry was
motivated by the need to accommodate the deficit of line response
exhibited by the recovered 1-d response functions $\Psi(\tau)$ for the
optical recombination lines on short timescales, by moving gas away
from the observers' line of sight, while at the same time reconciling
the measured delays to the hot dust ($\approx 50$ days) with
photoionisation model predictions of the radius at which the dust can
form ($\approx 100$ light-days for NGC~5548). Thus, as was first
suggested by Netzer and Laor (1993), the outer edge of the BLR is likely
largely determined by the distance at which grains can survive.  The
radially dependent scale-height is maintained by invoking a
macro-turbulent velocity field. Such a model is appealing not least
because AGN unification schemes which rely on orientation dependent
obscuration to distinguish between type {\sc i} and type {\sc ii}
objects, require viewing angles which peer over the edge of the
obscuring torus (ie. down into the bowl) for type {\sc i} objects.

In brief, the shape of the bowl is characterised in terms of its
scale height $H$,

\begin{equation}
H=\beta( r_{x})^{\alpha} \;  ,
\end{equation}

\noindent where $r_{x}$ is the projected radial distance along the
plane of the accretion disc (ie. $r_{x} = r \sin
\phi$, $r$ is the cloud source distance, $\phi$ is the angle
between the polar axis and the surface of the bowl), and $\alpha$,
$\beta$ are parameters which control the rate at which $H$ increases
with $r,\phi$. We adopt a circularised velocity field of the form

\begin{equation}
v_{\rm kep}^{2} = K \frac{ r_{x}^{2} } { (r_{x}^{2} + \beta^{2}r_{x}^{2\alpha})^{3/2} }
\end{equation}

\noindent where $v_{\rm kep}$ is the local Keplerian velocity and
$K=GM_{\rm BH}$, where $M_{\rm BH}$ is the mass of the black
hole. Using this formulism, when $\alpha=\beta=0$, the geometry
resembles a flattened disc with a standard Keplerian velocity field
$v_{kep}^{2} = GM_{\rm BH}/r_{x}$.  Here we adopt $\alpha=2$ and a
time-delay at the outer radius $\tau(r = R_{\rm out}) = (r - H)/c =
50$~days when viewed face-on, similar to the dust-delay reported for
the Seyfert 1 galaxy NGC~5548\footnote{The only measured delay for the
"hot" dust in the outer BLR for NGC~5548 (Suganuma et al. 2006) was
taken when NGC~5548 was in an historic low continuum flux
state. Taking the low-state source luminosity of NGC~5548 and an
appropriate ionising continuum shape, photoionisation models suggest
that silicate and graphite grains can form at radial distances of
$\approx$~50 days, consistent with the observations.}, yielding
$\beta=1/150$.  For the chosen black hole mass ($10^{8}$M$\odot$) and
continuum luminosity (representative of the mean ionising continuum
luminosity for the 13~yr monitoring campaign of NGC~5548), this model
BLR spans radial distances of 1.14--100 light-days.  We choose a
line-of-sight inclination $i=30$ degrees appropriate for the expected
inclination of type~{\sc i} objects, which results in a maximum delay at the
outer radius of 100 days for this geometry.  This we hereafter refer
to as the fiducial BLR geometry.

In the following simulations (\S3.2--\S3.4) we choose for simplicity a
radial surface emissivity distribution $F(r)$ for a generic
emission--line which is a power-law in radius, $F(r) \propto
r^{\gamma}$, with $\gamma$ values of $-1$, and $-2$, roughly spanning the
predicted range in power-law index for the radial surface emissivity
distribution of broad H$\beta$ over a range of BLR radii (see
e.g. Figure~2; Korista and Goad 2004; Goad, Korista and Ruff
2012). For a power-law radial surface emissivity distribution, the
radial responsivity distribution $\eta(r)=constant\; \forall r$ (thus
$\gamma$ values of $-1$, and $-2$ correspond to
$\eta(r)=-(\gamma/2)=0.5$ and $\eta(r)=-(\gamma/2)=1.0$ respectively).
{\bf In all cases we assume that the BLR boundaries remain fixed at
  their starting values}.

\subsection{The driving continuum light-curve}\label{drw}

We model the variable ionising UV continuum light-curve as a damped
random walk (DRW) in the logarithm of the flux, which has been shown
to be a good match to the observed continuum variability of quasars
(Uttley et al 2005; Kelly et~al. 2009; Kozlowski et~al. 2010; MacLeod
et~al. 2010; Czerny et~al. 2003; Zu et al. 2011). For a damped random
walk in the logarithm of the flux, the flux distribution function is
log-normal by construction.

To simulate the light-curve we first select a random variable with
mean $\mu$ and variance $\sigma^{2}T_{\rm char}/2$, where $T_{\rm
  char}$ is the characteristic timescale (often referred to as the
relaxation time or mean reversion timescale) of the DRW, and $\sigma$
represents the variability on timescales much shorter than $T_{\rm
  char}$.  For a given flux $X(s)$, the flux at a later time $t$,
$X(t)$ is then constructed piecewise, by selecting in turn a randomly
distributed variable with expectation

\begin{equation}
E <X(t)| X(s)> = e^{-\Delta t/T_{\rm char}}(X(s)-\mu) + \mu  \, ,
\end{equation}

\noindent and variance

\begin{equation}
Var <X(t)| X(s)>  = \frac{ \sigma^{2} T_{\rm char} }{ 2 } [1 - e^{-2 \Delta t/T_{\rm char}}]  \, .
\end{equation}

\noindent By setting $\Delta t=1$~day, the DRW becomes equivalent to an
AR(1) process (Kelly et al. 2009). The DRW is controlled by just 3
parameters, $\mu$, $\sigma$, and $T_{\rm char}$. Since in our
implementation $X(t), X(s)$ are logarithmic fluxes, $\sigma$ can be
written as $\sigma(\log X)$.

\begin{figure}
\resizebox{\hsize}{!}{\includegraphics[angle=0,width=8cm]{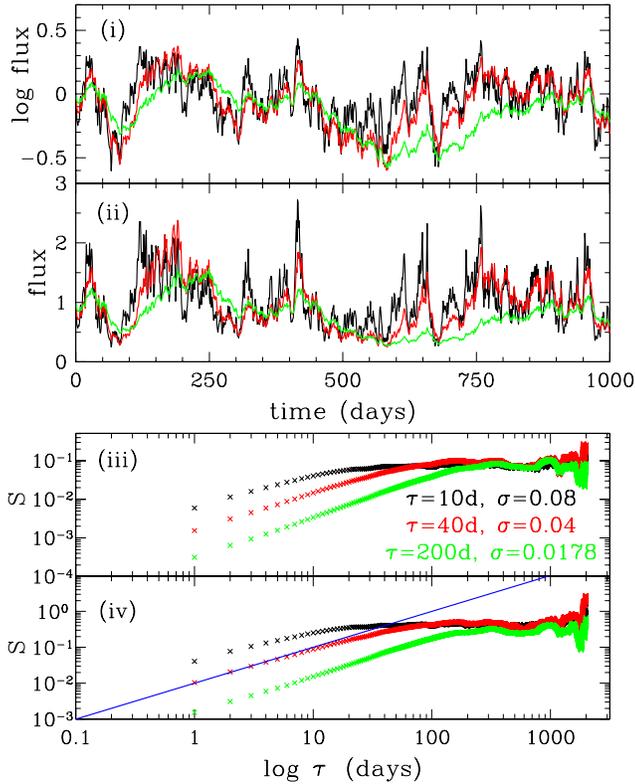}}
\caption{Example continuum light-curves. The light-curves are modelled
  as DRW in the logarithm of the flux (panel (i)) with constant
  long-term variance. Colours indicate characteristic timescales of
  $T_{\rm char}$ of 10 days (red), 40 days (red) and 200 days (green)
  respectively. Panel (ii) illustrates the transformed
  light-curves. Panel (iii) shows the structure function for the
  light-curves shown in panel (i).  Panel (iv) indicates the structure
  function for the transformed light-curves. The solid blue line
  indicates the expected power-law index for the structure function
  ($b=1$) appropriate for a Power Density Spectrum (PDS) $P(f) \propto
  f^{-\alpha}$, with $\alpha=2$, and here $f$ is the frequency.}
\label{plot_cont}
\end{figure}

Figure~\ref{plot_cont} panel (i) illustrates 3 DRWs with 1-day
sampling and characteristic timescales $T_{\rm char}$ of 10, 40 and
200~days. In each case $\sigma$ has been chosen to produce the same
long-term variance in each light-curve ($\sigma^{2} T_{\rm
  char}/2=0.032$). The transformed light-curves are shown in panel
(ii). In panels (iii) and (iv) we illustrate their respective
structure functions. In all cases, the structure functions asymptote
to twice the variance ($2\sigma^{2}$) on timescales of order the
characteristic timescale, $T_{\rm char}$, and to twice the variance of
the noise ($2\sigma_{n}^{2}$) on the shortest timescales (not shown).
 
Equipped with a method for constructing driving continuum light-curves
we are now in a position to drive our fiducial BLR geometry to
generate model emission-line light-curves and time-variable line
profiles from which we can measure the emission-line lag,
emission-line responsivity and the fwhm and dispersion of the
emission-line profile.

\subsubsection{The effect of light-curve duration and sampling frequency on the
measured line response}

As far as we are aware there are no extant studies into the effect of
light-curve duration and sampling frequency on measurement of the
effective emission-line responsivity, $\eta_{\rm eff}$. To remedy this
situation, we have performed detailed monte-carlo simulations to probe
the effect of campaign length and sampling frequency on the
determination of $\eta_{\rm eff}$. We employ two different techniques
for measurement of $\eta_{\rm eff}$, thereby allowing us to compare
their relative robustness to light-curve duration and sampling
patterns.

We determine the emission-line light curve by driving our fiducial BLR
model as described in section \S3.1 with a simulated DRW (\S\ref{drw})
in the logarithm of the flux assuming a locally linear response
approximation. Thus for these simulations
measured differences in the emission-line response (e.g. mean delay,
and line responsivity $\eta_{\rm eff}$ are governed only by
differences in the characteristics of the driving continuum
light-curve ($\sigma$, $T_{\rm char}$).

Unless otherwise specified we adopt a characteristic continuum
variability timescale $T_{\rm char}=40$ days, similar to that measured
for the UV continuum in the well-studied Seyfert 1 galaxy NGC~5548
(Collier et al. 2001) and $\sigma=0.04$, giving a long-term variance
$\sigma^{2} T_{\rm char}/2$ = 0.032.

\subsubsection{Duration}\label{tdur}

First we examine the effect of performing different duration observing
campaigns on the measured emission-line responsivity, assuming uniform
1-day sampling for input continuum light-curves spanning durations of
$T_{\rm dur}= 100$, 200, 500, 1000 and 1500~days.  For each input
continuum and resultant emission-line light-curve combination, we add
a random noise component drawn from a random Gaussian deviate with
$\sigma = 0.01*f$, where $f$ is the flux. Each point is then assigned
an error bar in a similar fashion.

\begin{figure}
\resizebox{\hsize}{!}{\includegraphics[angle=0,width=8cm]{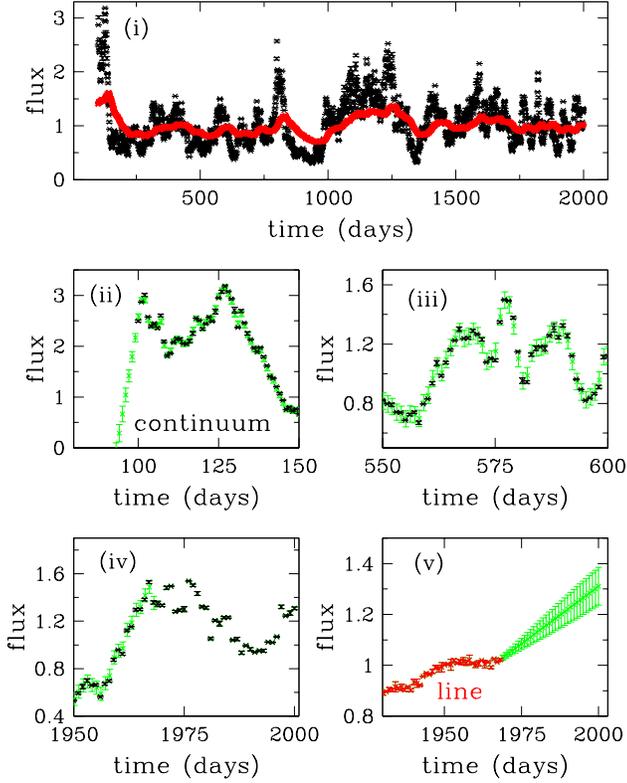}}
\caption{Panel (i) -- model continuum (black) and emission-line (red)
  light-curves.  Panels (ii)--(iv) - original continuum data points
  (black) and estimated continuum values (green points) corresponding
  to the lag-corrected emission-line epochs with values determined by
  interpolating between the bracketing continuum points and errors
  estimated from the structure function of the continuum
  light-curve. Panel (ii) indicates the extrapolation of the continuum
  at shifted line epochs prior to the start of the campaign. Panels
  (iii) and (iv) indicate interpolated continuum points (green).
  Panel (v) shows the extrapolated line light-curve (green points) for
  those continuum epochs at the end of the campaign (panel (iv)) for
  which there are no corresponding line data. Here we have assumed a
  power-law radial emissivity law $F(r) \propto r^{\gamma}$, with
  $\gamma=-1$, and $\eta_{\rm eff} = -(\gamma/2)=0.5$ }
\label{prep}
\end{figure}

We calculate the effective emission-line responsivity
$\eta_{\rm eff}$ using the two methods described in \S2, thereby
allowing us to compare their relative merits.  The simplest method for
calculating $\eta_{\rm eff}$ utilises the ratio of the fractional
variance in the emission-line relative to the fractional variance in
the continuum (see \S2.2.1). For each continuum--emission-line light-curve
combination we determine $\eta_{\rm eff} = F_{\rm var}(\rm
line)/F_{\rm var}(\rm cont)$, and its dispersion $\sigma$, from the
centroid and dispersion of the distribution function of $\eta_{\rm
  eff}$ computed from bootstrap resampling of the input continuum and
emission-line light-curves (using 10,000 trials with full
replacement).  Note that for real data, each light-curve should first
be corrected for contaminating non-varying components, for example the
host galaxy contribution to the continuum light-curve, and the narrow
emission-line contribution to the broad emission-line fluxes. While
the narrow-line contribution to the broad emission-line flux is
relatively straightforward to measure, particularly if low-state
spectra exist, contributions from stellar light in the host-galaxy to
the measured continuum flux are generally more problematic (Bentz et
al. 2006). However, host galaxy contamination of the continuum
light-curve can be mitigated if UV continuum measurements are
available, since the contribution of stellar light to the UV continuum
is modest at best.

\begin{figure}
\resizebox{\hsize}{!}{\includegraphics[angle=0,width=8cm]{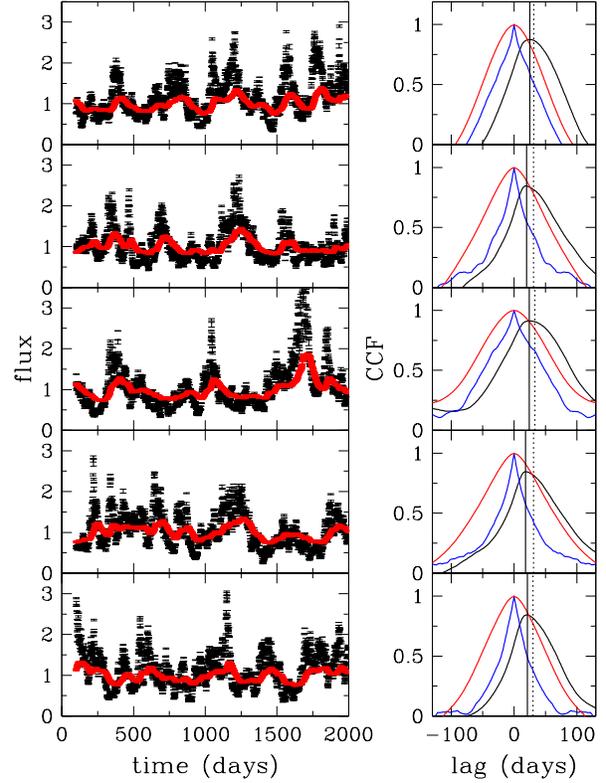}}
\caption{Left-hand panels - Example continuum (black points) and model
  emission-line light-curves (red points), see text for
  details. Right-hand panels - the corresponding cross-correlation
  function (black line), continuum auto-correlation function (blue
  line) and line auto-correlation function (red line). The vertical
  solid line indicates the CCF peak (or lag), the vertical dashed line
  the CCF centroid. Note that the measured lag (CCF peak of centroid)
  is sensitive to the form of the driving continuum light-curve, as
  denoted by the continuum ACF. Here we have assumed a power-law radial emissivity law  $F(r) \propto r^{\gamma}$, with
  $\gamma=-1$, and $\eta_{\rm eff} = -(\gamma/2)=0.5$.}
\label{ccf}
\end{figure}

For the second method, we adopt the approach used in Goad, Korista and
Knigge (2004), using $d \log f_{\rm line}/d \log f_{\rm cont}$ to
determine $\eta_{\rm eff}$, having first corrected for contaminating
narrow-line contributions to the broad emission-line flux and the host
galaxy contribution to the continuum flux, and the ``average delay''
between the continuum and emission-line light-curves, using their
respective structure functions to estimate the uncertainties on
interpolated and extrapolated points (see Figure~\ref{prep}).  Thus
for each continuum--emission-line light-curve combination we first
compute the centroid of the cross-correlation function (hereafter,
CCF) between the continuum and emission-line light-curve, measured
above a CCF threshold of 0.6 of the peak correlation\footnote{In a
  blind search designed to find the peak of the CCF between
  pre-specified minimum and maximum delays, not all trials will result
  in CCFs which meet this criteria, particularly for short light-curve
  durations (ie. for light-curve durations shorter than the
  autocorrelation function of the continuum light-curve), or for
  light-curve segments which show little variation.}. Example model
continuum and emission-line light-curves and their corresponding CCFs
are shown in Figure~\ref{ccf} for an assumed power-law radial surface
emissivity distribution $F(r)\propto r^{\gamma}$, with $\gamma=-1$,
corresponding to $\eta_{\rm eff} = -(\gamma/2)=0.5$ (e.g. as indicated
by the green lines in Figure~2). Note that even for a characteristic
damping timescale of 40~days for the driving UV continuum, the form of
the continuum ACF varies considerably, depending on the nature of the
continuum variability over the campaign duration.
\begin{figure}
\resizebox{\hsize}{!}{\includegraphics[angle=0,width=8cm]{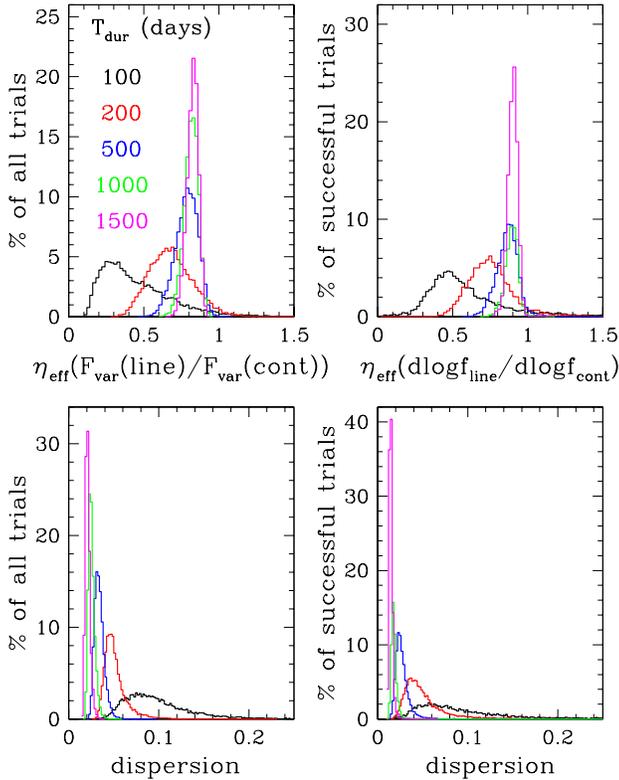}}
\caption{Upper panels - the distribution function for $\eta_{\rm eff}$
  as measured using the ratio of the variance in the line relative to
  the variance in the continuum (upper left), and from the ratio 
$d\log f_{\rm line} /d\log f_{\rm cont}$ after first correcting for
  the delay between the continuum and emission-line variations (see
  text for details). Colours represent results from simulated
  light-curves with durations of 100 (black), 200 (red), 500 (blue),
  1000 (green), and 1500 (magenta) days, each sampled at 1-day
  intervals. Lower panels -- the corresponding distribution function
  for the dispersion on $\eta_{\rm eff}$ from each of the simulated
  light-curves measured using bootstrap resampling of the light-curves
  with full replacement. Here we have assumed a power-law radial
  surface emissivity distribution $F(r) \propto r^{\gamma}$, with
  $\gamma=-2$, and $\eta_{\rm eff} = -(\gamma/2)=1$.}
\label{f_tdur}
\end{figure}

\begin{figure}
\resizebox{\hsize}{!}{\includegraphics[angle=0,width=8cm]{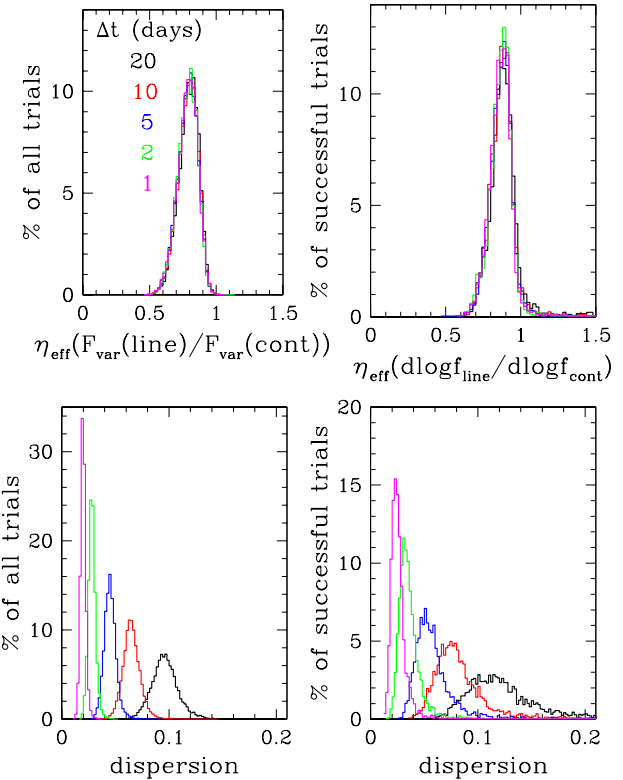}}
\caption{Effect of the sampling pattern on the measured response
  $\eta_{\rm eff}$. For a fixed light-curve durations of 500 days, we
  resample the continuum and emission-line light-curves a fixed number
  of times. The distribution in $\eta_{\rm eff}$ as determined from measurements
  of the ratio of the fractional variation in the line relative to the
  fraction variation in the continuum for 10,000 simulated
  light--curves, is unaffected by changes in the sampling rate, for
  fixed duration light-curve (upper left-panel). However, the
  dispersion in the measured slope determined from bootstrap
  resampling, increases significantly at lower sampling rates. Here we
  have assumed a power-law radial surface emissivity distribution $F(r)
  \propto r^{\gamma}$, with $\gamma=-2$, and $\eta_{\rm eff} =
  -(\gamma/2)=1$.}
\label{sampw}
\end{figure}
We then shift the emission-line light-curve by the measured delay (the
CCF centroid), and reconstruct the continuum light-curve at the
corresponding times by interpolation, with flux uncertainties
determined directly from the structure function. After applying the
shift to the emission-line light curve, there will be no continuum
points corresponding to the first few line points and conversely no
emission-line measurements corresponding to the last few continuum
points. These are determined by extrapolation with uncertainties
determined from their respective structure functions.  In our
implementation, we construct structure functions for both line and
continuum light-curves, which are then smoothed using a Savitzky-Golay
filter (Press et al. 1992), this effectively smooths the data without
any loss of information.

The effective line responsivity, $\eta_{\rm eff}$, the power-law index
relating the continuum and emission-line luminosities (equation 1),
the quantity of most interest to observers, is determined from
least-squares fitting with errors in both $x$ and $y$, with an error
estimate (1$\sigma$ uncertainty) in individual slope determinations
determined from bootstrap resampling of the original data with full
replacement (10,000 trials).  Thus for each continuum--emission-line
light curve combination we have a single estimate of $\eta_{\rm eff}$
and its associated uncertainty $\sigma$.  The whole procedure is then
repeated for 10,000 realisations of the driving continuum
light-curve. From these we can compute the distribution functions for
$\eta_{\rm eff}$ and $\sigma$.  The results of our simulations are
shown in Figure~\ref{f_tdur}, for (i) $\eta_{\rm eff}$ as determined
from the ratio of the fractional variance in the line relative to the
fractional variance in the continuum (upper left panel), and (ii)
$\eta_{\rm eff}$ as determined from $d\log f_{\rm
line}/d\log f_{\rm cont}$ (upper right panel). The lower
panels indicate the corresponding distribution functions for the
dispersion on the individual measurements, determined from the
bootstrap method.  The distribution functions only indicate successful
trials, which for the second method depends on whether or not
measurement of the CCF centroid meets the required threshold of 0.6
between the pre-specified minimum and maximum delays\footnote{For
  method 2 the centroid of the CCF can be found in $\approx$80\% of
  all trials for the shortest duration light-curves. We could of
  course have used the peak of the CCF (or lag) for shifting the data,
  giving a 100\% success rate, but chose not to do so because of the
  inherent bias toward smaller delays that this would introduce.}.

Our simulations indicate that both methods used in the determination
of $\eta_{\rm eff}$ yield similar results, suggesting that either
method may be employed to estimate $\eta_{\rm eff}$ from observational
data (Figure~\ref{f_tdur}, upper panels). However, Figure~\ref{f_tdur}
also indicates that even with a 1500~day light-curve with daily
sampling, the measured responsivity $\eta_{\rm eff}$, with a range of
$0.7 < \eta_{\rm eff} < 1$ is already less than the input value.  This
we attribute to geometric dilution of the continuum variations by the
spatially extended BLR. We explore this further in \S\ref{dilution}.
Moreover, for our chosen BLR model and driving continuum light-curve
large variations in $\eta_{\rm eff}$ can be found for light-curve
durations shorter than 500 days.  In particular, for the light-curve
durations of $\sim$100 days, the distribution in $\eta_{\rm eff}$ is
both significantly displaced (factor of 2 or more) from the expected
value (in the absence of geometric dilution $\eta_{\rm eff}=1.0$ for
power-law radial surface emissivity distribution of slope
$\gamma=-2$), and becomes highly skewed, with an extended tail toward
large $\eta_{\rm eff}$ values. In addition the typical uncertainty on
any single measurement also increases as the campaign duration
decreases (lower panels). This suggests that campaign durations of a
factor of a few longer than the width of the continuum
auto-correlation function (ACF) are required in order to remove the
influence of windowing effects on the measured value of $\eta_{\rm
eff}$ and reduce its uncertainty\footnote{We note in passing that
Welsh (1999) reported that, in a similar fashion, the bias toward
smaller delays and large variance in recovered lags are also a
consequence of finite duration sampling and the dominance of long
timescale trends in the light curves, and not due to noise or
irregular sampling.}. Collier et~al. (2001) and Horne et~al. (2002)
showed that accurate recovery of the emission-line response function
also requires high cadence long duration monitoring campaigns.
We note that if instead we had used a substantially smaller (more
compact) BLR, then the impact of campaign duration on the measured responsivity
$\eta_{\rm eff}$ and delay $\tau$ would be significantly reduced,
particularly for short campaign durations.

We note that for the models described here $\eta_{\rm eff}$ will not
in general correlate with continuum level. This arises because for
simple power-law radial surface emissivity distributions $F(r)\propto
r^{\gamma}$, and therefore $\eta(r)=-(\gamma/2)=constant\; \forall r$, while the
boundaries of our model BLR are fixed both spatially and
temporally. For our chosen driving continuum, the mean ratio between
the maximum and minimum flux is $10 \pm 3$ (based on 10,000 simulated
light-curves), similar to the range measured for the optical continuum
light-curve of NGC~5548 during the 13~yr ground-based monitoring
campaign of this source.

\subsubsection{Effect of sampling window}\label{sw}

Observations from the ground rarely (if ever) approach regular
sampling. To illustrate the effect of the sampling pattern on the
measured responsivity, we adopt a 500 day light-curve as a baseline. For
each 500 day continuum and emission-line light-curve we randomly
select an integer number of points with replacement, the rest are then
discarded. We then follow the procedure above to determine $\eta_{\rm eff}$.
 We aim for mean sampling rates of 1,2, 5, 10 and 20
days (equivalent to 500, 250, 100, 50, and 25 points
respectively). Figure~\ref{sampw} shows the results of these
simulations. Reducing the sampling rate does not significantly alter
the distribution function for $\eta_{\rm eff}$. However, the mean
dispersion on an individual measurement, increases significantly as
the number of sampling points is reduced (Figure~\ref{sampw} lower
panel). Note that the dispersion on a single measurement (from
bootstrap resampling) is smaller than the dispersion on $\eta_{\rm
eff}$ from different realisations of the same process.  Thus our
simulations demonstrate that both methods of measuring $\eta_{\rm
eff}$ are robust to low sampling rates, though the error in an
individual measurement will increase as the sampling rate decreases.

\begin{figure}
\resizebox{\hsize}{!}{\includegraphics[angle=0,width=8cm]{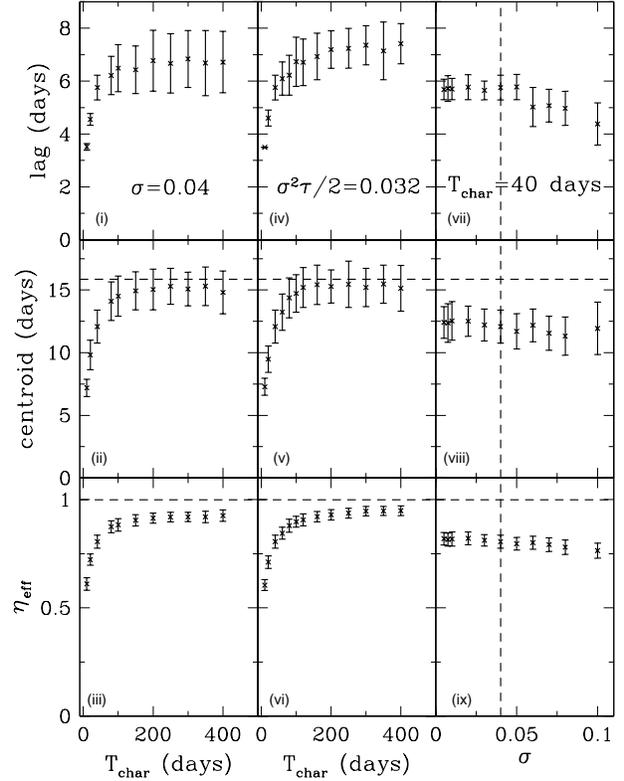}}
\caption{Panels (i)--(iii) -- the variation in CCF lag, CCF centroid
  and emission-line responsivity $\eta_{\rm eff}$ as a function of
  characteristic timescale $T_{\rm char}$ for driving continuum
  light-curves with a fixed short-timescale variability amplitude
  $\sigma=0.04$.  Panels (iv)--(vi) -- the variation in CCF lag, CCF
  centroid and emission-line responsivity $\eta_{\rm eff}$ as a
  function of $T_{\rm char}$ for fixed long-term variance
  $\sigma^{2}T_{\rm char}/2$. Panels (vii)--(ix) -- the variation in
  CCF lag, CCF centroid and emission-line responsivity $\eta_{\rm
    eff}$ for fixed $T_{\rm char}$ and varying $\sigma$.  The dashed
  horizontal lines in panels (ii), (v) and (viii) represent the
  centroid of the 1-d response function.  The dashed horizontal lines
  in panels (iii),(vi) and (ix) indicate the expected responsivity in
  the absence of geometric dilution. The dashed vertical line in
  panels (vii), (viii) and (ix) represents a plausible upper limit to the
  short-timescale variability amplitude. In all cases we assume $F(r)
  \propto r^{\gamma}$, with $\gamma=-2$, and $\eta_{\rm eff} =
  -(\gamma/2)=1$.}
\label{plot_kelly}
\end{figure}

In summary, provided that the sampling rate is irregular, so that both
low and high frequency variations are adequately sampled, the measured
line responsivity $\eta_{\rm eff}$` is most sensitive to the duration
of the observing campaign. In particular, for campaigns which are
shorter than a few times the light crossing time of the outer boundary
of our fiducial BLR ($\sim$100 days), $\eta_{\rm eff}$ can deviate
significantly from its expected value of 1.0, being generally biased
toward smaller values. This result can be used to inform the design of
future reverberation mapping experiments of individual sources.

\subsubsection{The dependence of emission-line lag and emission-line responsivity on $T_{\rm char}$, $\sigma$}

To test the sensitivity of our measurements of the CCF peak, CCF
centroid and line responsivity $\eta_{\rm eff}$ to those parameters
which control the short- and long-term variability of the driving
continuum light-curve (namely $\sigma$, $T_{\rm char}$) we have
simulated sets of fake continuum light-curves, with (i) constant
$\sigma$, (ii) constant $T_{\rm char}$, and (iii) constant long-term
variance ($\sigma^{2} T_{\rm char}/2$), and assuming fixed BLR
boundaries.  Our choice of power-law radial surface emissivities and
fixed BLR boundaries allows us to cleanly isolate the effect of these
parameters on the measured lag and responsivity, from those effects
associated with ``breathing'', since under these conditions no
breathing can occur.

For models (i) and (iii) we vary $T_{\rm char}$ from 10--400 days,
encompassing the range in characteristic timescales for the continuum
variability of a typical Seyfert 1 galaxy\footnote{Kelly~et al. (2009)
  measured a characteristic timescale of $\approx 200$ days for
  NGC~5548 using the 13 yrs of ground-based optical monitoring data
  obtained by AGN Watch. This is significantly smaller than the
  estimate of $\approx 1000$~days reported by Czerny et al. 2003, and
  a factor of a few longer than the value of 40~days reported for the
  UV continuum variability timescale by Collier et al. (2001)
  determined using a structure function analysis of the UV continuum
  as observed during the 1989 IUE monitoring campaign on this
  source.}. At each value we generate 100 simulated light-curves
sampled at 1~day intervals and with a duration of $T_{\rm
  dur}=2000$~days, more than long enough to avoid sampling and
windowing effects (\S\ref{tdur}--\S\ref{sw}). We then drive our
fiducial BLR model, and generate a corresponding emission-line
light-curve. We add a random Gaussian deviate to simulate noise and
assign a random error again drawn from Gaussian distribution with
dispersion $\sigma$ equivalent to 1\% of the flux at that epoch.  For
each continuum--emission-line light curve pair we then estimate the
CCF peak and CCF centroid (for the latter using a threshold of 60\% of
the CCF peak), using the implementation of Gaskell and Peterson
(1987), interpolating in both light-curves. The emission-line
responsivity $\eta_{\rm eff}$ we estimate from the ratio of the
fractional variation in the line to the fractional variation in the
continuum (\S2), with an error estimate determined from bootstrap
resampling. The results of this analysis are shown in
Figure~\ref{plot_kelly}.  Here, the points and their error bars
represent the centroid and 1$\sigma$ dispersion of the distribution in
values determined from 100 simulated light-curves.

For constant $\sigma$ and $T_{\rm char} \ge 100$~days (the maximum
delay at the outer radius for our fiducial BLR geometry when viewed at
an inclination of 30 degrees), both the CCF peak, CCF centroid and the
maximum of the continuum--emission line cross-correlation coefficient
are constant (Figure~\ref{plot_kelly} panels (i)--(ii)) and decline
sharply for smaller $T_{\rm char}$. Similarly, the effective line
responsivity $\eta_{\rm eff}$, is also constant for large $T_{\rm
  char}$, though slightly smaller than the expected value of 1.0 in
the absence of geometric dilution for our chosen emissivity law, and
again declines sharply as $T_{\rm char}$ is reduced (see e.g. panel
(iii)). Panels (iv)--(vi) of Figure~\ref{plot_kelly} show the
variation in each of the measured parameters for fixed long-term
variance. As with panels (i)--(iii), each of the parameters is
constant for characteristic timescales longer than 100~days and
declines rapidly for timescales that are shorter than this.  Panels
(vii)--(ix) of Figure~\ref{plot_kelly} suggest that the variation in
the CCF peak , CCF centroid and line responsivity $\eta_{\rm eff}$ are
independent of $\sigma$ for small $\sigma$ ($\sigma$ values larger
than $\sigma \sim 0.06$ produce driving continuum light-curves which
undergo large amplitude short-timescale variations which are likely
unphysical).

Thus in the absence of windowing effects (ie. for light-curves of
significantly long duration), and for a simple power-law radial
surface emissivity distribution, {\em the measured line responsivity
$\eta_{\rm eff}$ and delay are determined by the characteristic
timescale of the driving continuum light-curve $T_{\rm char}$ in
relation to the the maximum delay at the outer radius for a particular
line, $\tau_{\rm max}$(line), given the geometry and observer line of
sight inclination\/}.  If $T_{\rm char}$ is small compared to
$\tau_{\rm max}$(line), then the continuum signal is significantly diluted
(geometric dilution), resulting in small $\eta_{\rm eff}$ and smaller
delays (e.g. Figure~9, panels (i)--(vi)). Conversely, geometric
dilution is minimised if $T_{\rm char} \ge \tau_{\rm max}$(line).

Note that if we had instead chosen a flatter emissivity distribution
(e.g. $\gamma=-1$), the measured responsivity would also be reduced
because (i) locally the responsivity is smaller ($\eta(r) =
-(\gamma/2)=0.5$), and (ii) the larger responsivity-weighted radius
resulting from the flatter emissivity distribution would reduce the
coherence of the continuum--emission-line variations resulting in
larger CCF centroids and lags. Thus for a flatter radial surface
emissivity distribution, the knee in the relation shown in panels
(iii) and (vi) of Figure~9, would move down and towards the right.
Again, for $T_{\rm char} \ge \tau_{\rm max}$(line) geometric dilution
is minimised.

Figure~\ref{plot_kelly} indicates that the dependence of the CCF
centroid, CCF lag and $\eta_{\rm eff}$ on $T_{\rm char}$ for $T_{\rm
  char} \le \tau_{\rm max}$(line) are broadly similar (cf. panels
(i)--(iii)). If we take the measured CCF centroid and CCF lag and plot
them against $\eta_{\rm eff}$ (for a fixed $T_{\rm char}$), we find
that they follow a simple linear relation of the form $\eta_{\rm eff}
= m \tau + k$, with coefficients $m=0.038$, $k=0.343$ and $m=0.0948$,
$k= 0.280$ for the CCF centroid and lag respectively. This suggests
that $T_{\rm char}$ determines the extent to which a BLR of fixed
radial extent is probed by the continuum variations, and that this in
turn sets the value of $\eta_{\rm eff}$.  We expand upon the impact of
a particular choice of BLR geometry and BLR size in determining the
measured responsivity and delay for fixed $T_{\rm char}$ in the
following section.

As a general point of interest, we note that if among the general AGN
population there exists, for a fixed continuum luminosity, a range in
BLR sizes and characteristic continuum variability timescales $T_{\rm
  char}$ then the varying degrees of geometric dilution exhibited by
the AGN population will introduce significant scatter into the BLR
size--luminosity relation.

\begin{figure}
\resizebox{\hsize}{!}{\includegraphics[angle=0,width=8cm]{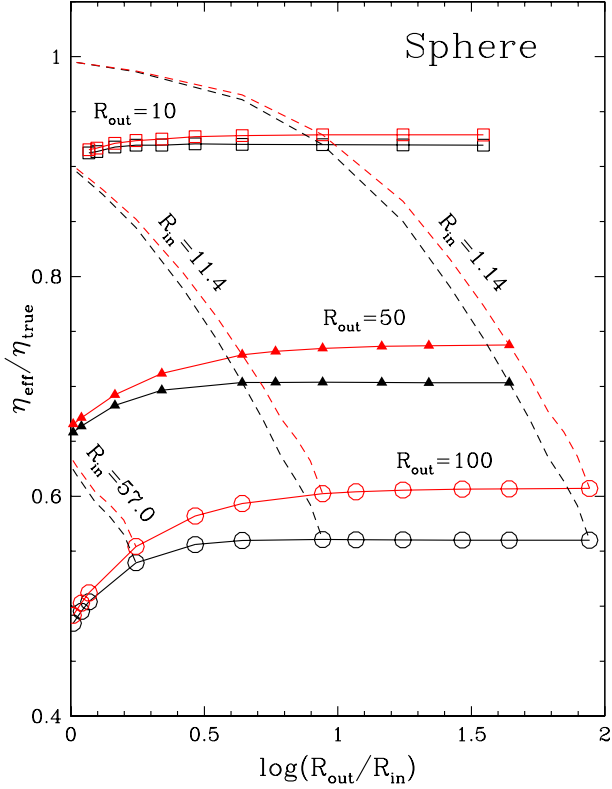}}
\caption{$\eta_{\rm eff}/\eta_{\rm true}$ as a function of $R_{\rm
    out}/R_{\rm in}$ for a spherical geometry and fixed $R_{\rm out}$
  of 100 (open circles), 50 (filled triangles) and 10 (open squares)
  light-days. Also shown (dashed lines), are $\eta_{\rm eff}/\eta_{\rm
    true}$ as a function of $R_{\rm out}/R_{\rm in}$ for fixed $R_{\rm
    in}$ of 1.14, 11.4 and 57.0 light-days. The colours refer to the
  slope of the radial power-law emissivity distribution (black:
  power-law slope $-1$, $\eta(r)=-(\gamma/2)=0.5$, red: power-law
  slope $-2$, $\eta(r)=-(\gamma/2)=1.0$).}
\label{eta_sphere}
\end{figure}

\begin{figure}
\resizebox{\hsize}{!}{\includegraphics[angle=0,width=8cm]{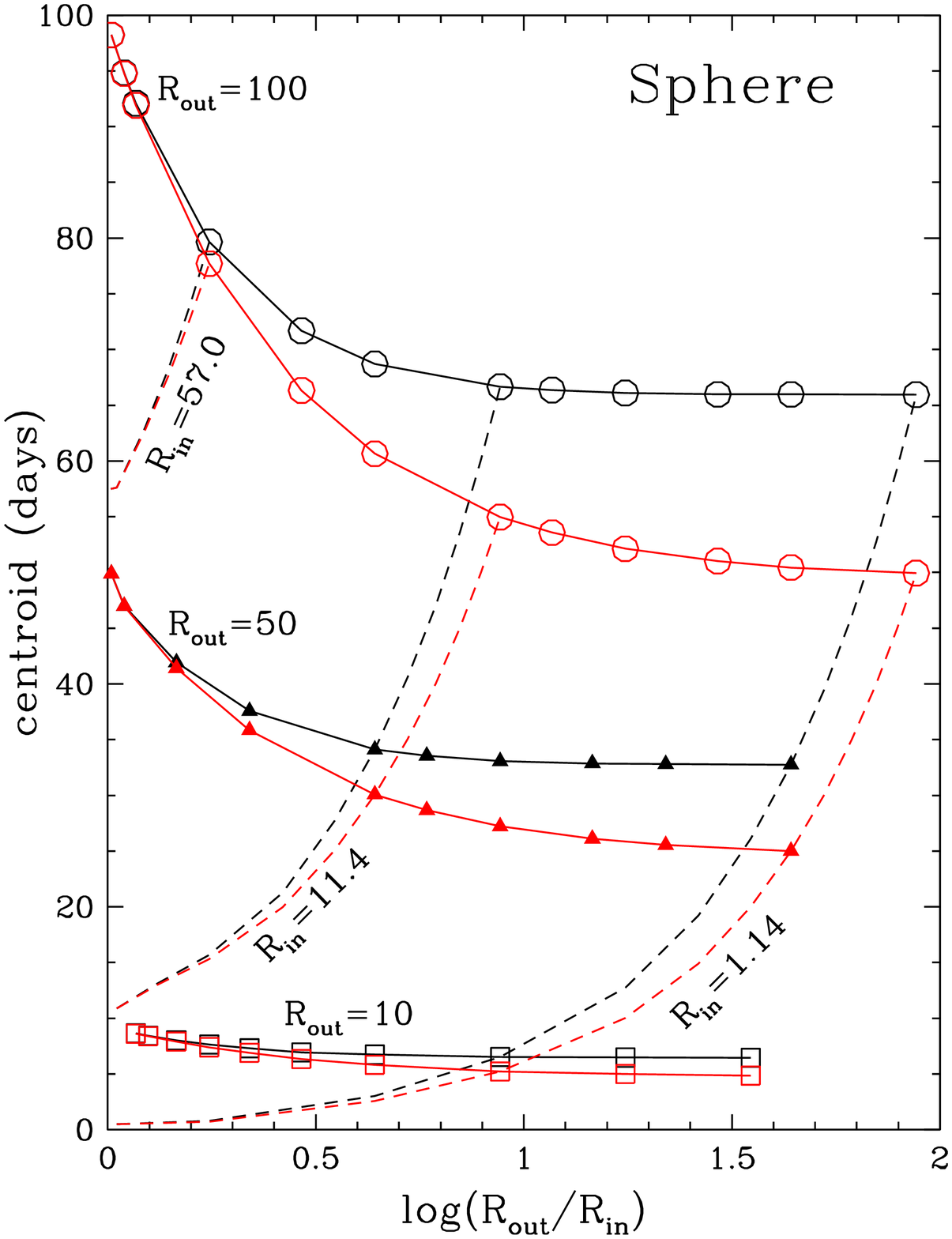}}
\caption{The centroid of the 1-d response function as a function of
$R_{\rm out}/R_{\rm in}$ for a spherical geometry and fixed $R_{\rm
out}$ of 100 (open circles), 50 (filled triangles) and 10 (open
squares) light-days. Also shown (dashed lines) is the centroid of the
1-d response function as a function of $R_{\rm out}/R_{\rm in}$ for
fixed $R_{\rm in}$ of 1.14, 11.4 and 57.0 light-days. The colours
refer to the slope of the radial power-law emissivity distribution
(black: power-law slope $-1$, $\eta(r)=-(\gamma/2)=0.5$, red:
power-law slope $-2$, $\eta(r)=-(\gamma/2)=1.0$).}
\label{lag_sphere}
\end{figure}

\begin{figure}
\resizebox{\hsize}{!}{\includegraphics[angle=0,width=8cm]{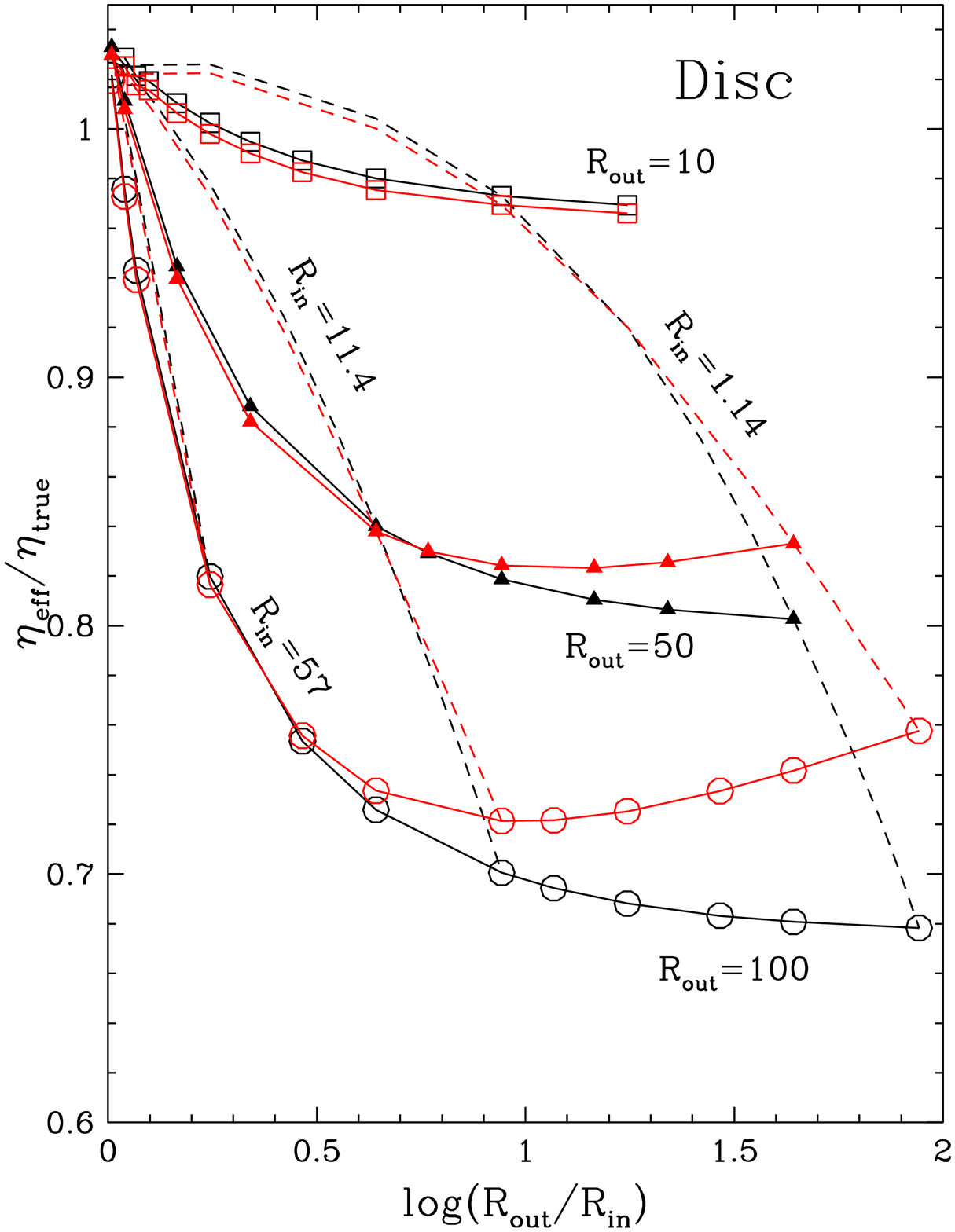}}
\caption{$\eta_{\rm eff}/\eta_{\rm true}$ as a function of
$R_{\rm out}/R_{\rm in}$ for a thin-disc geometry and fixed $R_{\rm
out}$ of 100 (open circles), 50 (filled triangles) and 10 (open
squares) light-days. Also shown (dashed lines), are $\eta_{\rm
eff}/\eta_{\rm true}$ as a function of $R_{\rm out}/R_{\rm in}$ for
fixed $R_{\rm in}$ of 1.14, 11.4 and 57.0 light-days. The colours
refer to the slope of the radial power-law emissivity distribution
(black: power-law slope $-1$, $\eta(r)=-(\gamma/2) = 0.5$, red: power-law slope $-2$,
$\eta(r)=-(\gamma/2)=1.0$).}
\label{eta_disk}
\end{figure}

\begin{figure}
\resizebox{\hsize}{!}{\includegraphics[angle=0,width=8cm]{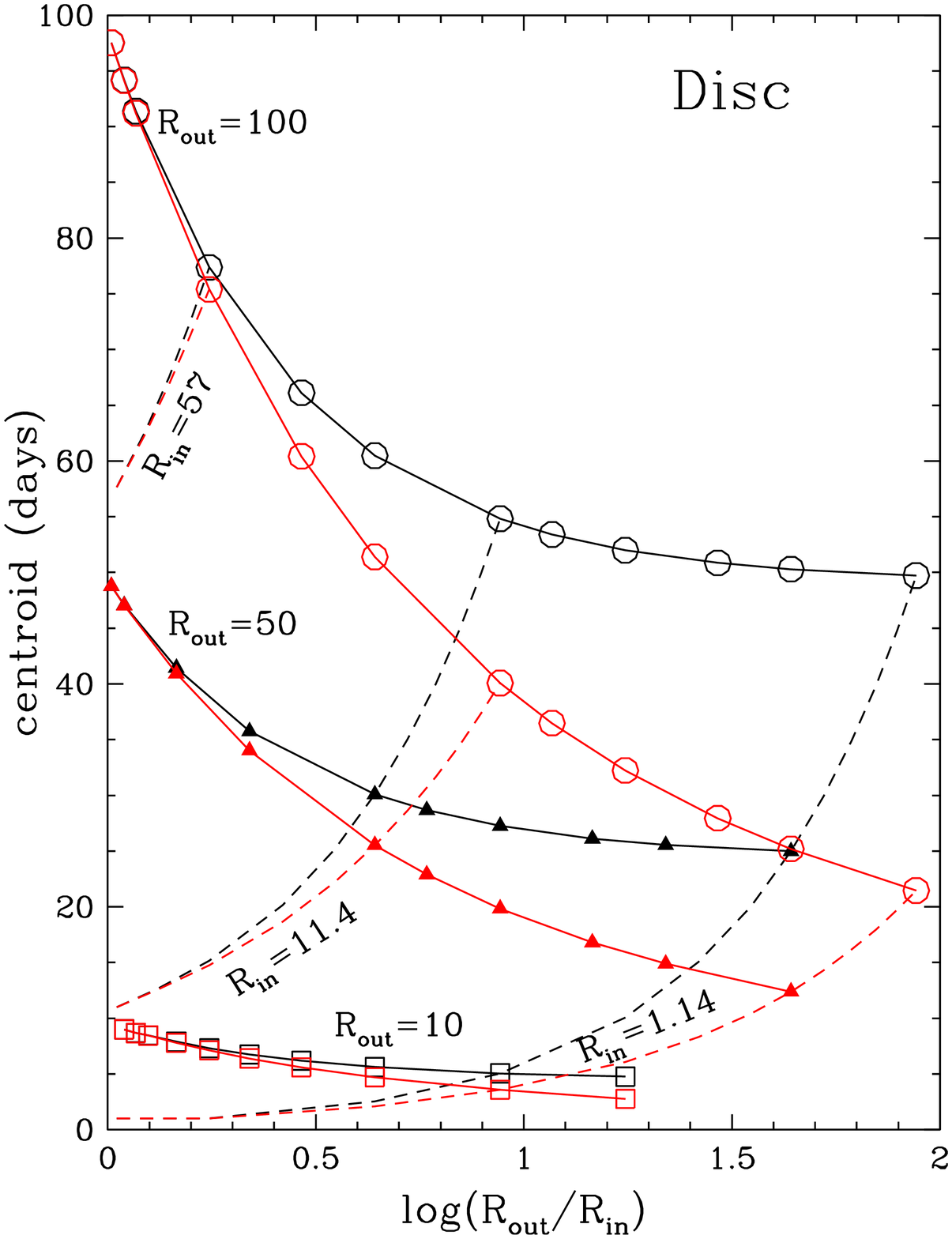}}
\caption{The centroid of the 1-d response function as a function of
$R_{\rm out}/R_{\rm in}$ for a disc-shaped BLR geometry and fixed
$R_{\rm out}$ of 100 (open circles), 50 (filled triangles) and 10
(open squares) light-days. Also shown (dashed lines) is the centroid
of the 1-d response function as a function of $R_{\rm out}/R_{\rm in}$
for fixed $R_{\rm in}$ of 1.14, 11.4 and 57.0 light-days. The colours
refer to the slope of the radial power-law emissivity distribution
(black: power-law slope $-1$, $\eta(r)=-(\gamma/2)=0.5$, red:
power-law slope $-2$, $\eta(r)=-(\gamma/2)=1.0$).}
\label{lag_disc}
\end{figure}

\subsection{The role of geometric dilution}\label{dilution}

As noted above, for BLR geometries with significant spatial extent
relative to the characteristic timescale of the driving continuum, we
expect a degree of geometric dilution of the measured line
responsivity.  In previous work, Pogge and Peterson (1992) suggested
that a first order correction for the continuum emission-line delays
is sufficient to allow an accurate determination of $\eta_{\rm eff}$,
showing that data corrected in this fashion displays significantly
reduced scatter. Following on from this, Gilbert and Peterson (2003)
demonstrated that the $\eta_{\rm eff}$ recovered from the data depends
critically upon the fitting process employed, finding values of
between 0.53--0.65 for broad H$\beta$ over the full 13~yr ground-based
optical monitoring campaign of NGC~5548.  They also investigated the
dependence of the line responsivity on BLR size, using as an example
geometrically-thin spherical BLR geometries represented by top-hat
transfer functions. They found that $\eta_{\rm eff}$ varies by only
$\approx$5\% between transfer functions of half-width 1- and 10 days,
so that while the line responsivity decreases as the BLR outer
boundary increases, it does so at a very slow rate. Here we expand on
this work by examining the role of geometric dilution for spatially
extended BLR geometries, and for which a top-hat transfer function is
no longer appropriate.

We start by generating spherical, disc, and bowl-shaped BLR
geometries, specified by an inner radius $R_{\rm in}$, outer radius
$R_{\rm out}$, and for which the emission-line response is taken to be
constant across the whole BLR (ie. a locally-linear response
approximation ($\eta(r)=constant$ $\forall$ $r$). Initially, both the
disc and bowl-shaped BLRs are assumed to be viewed face-on ($i=0$) by
an external observer. For non-spherical BLR geometries, other values
of observer line of sight inclination only serve to increase the
effect of geometric dilution, since for these geometries the spread in
delays is orientation dependent and is a maximum when viewed
edge-on.

Values of $\eta(r)$ were chosen to encompass the range of values
measured in standard photoionisation model calculations of the BLR
gas, adopting power-law radial surface line emissivity distributions
($F(r) \propto r^{\gamma}$) with power-law indices of $-1$ and $-2$
(equivalently $\eta(r)=0.5$, 1.0 respectively). For each model BLR, we
drive the emission-line response with a continuum light-curve modelled
as a damped random walk (Kelly et al. 2009; MacLeod et al. 2010) with
{\bf a characteristic continuum variability time-scale $T_{\rm char}=
  40$~days}, appropriate for the UV continuum variability observed in
the nearby Seyfert 1 galaxy NGC~5548 (Collier and Peterson 2001), and
assuming stationary BLR boundaries. This particular choice of $T_{\rm
  char}$ ensures that the continuum variability timescale is
well-matched to the responsivity-weighted size of the BLR over the
range in BLR spatial extent explored here.  From the input continuum
light-curve and resultant broad emission-line light-curve we compute
the ratio $\eta_{\rm effl}/\eta_{\rm true}$, where $\eta_{\rm true}$
is the expected responsivity in the absence of geometric dilution, and
$\eta_{\rm eff}$ is the measured effective responsivity as determined
from the ratio of the fractional variability of the line relative to
the fractional variability of the continuum ($F_{\rm var}(\rm
line)/F_{\rm var}(\rm cont)$).  We have verified using monte carlo
simulations that for the duration of the input continuum light-curve
used here, the ratio $F_{\rm var}(\rm line)/F_{\rm var}(\rm cont)$ is
not sensitive to windowing effects (the dispersion $\sigma$ on
$\eta_{\rm eff}$ is less than 0.02 for 1-day sampling over 1000 days
duration, see e.g. Figure~\ref{f_tdur}).

\subsubsection{Spherical BLR geometries}

Figure~\ref{eta_sphere} indicates the ratio $\eta_{\rm eff}/\eta_{\rm
  true}$ for isotropically emitting BLR clouds occupying a spherical
BLR with a range of $R_{\rm out}/R_{\rm in}$, and assuming fixed
$R_{\rm out}$ of 100 (open circles), 50 (filled triangles) and 10
(open squares) light-days. Similarly the dashed lines (running from
top--bottom) indicate $\eta_{\rm effl}/\eta_{\rm true}$ as a function
of $R_{\rm out}/R_{\rm in}$, for fixed $R_{in}$. The colours refer to
the power-law index $\gamma$ of the radial surface emissivity
distribution (black: $\gamma=-1$, $\eta=0.5$, red: $\gamma=-2$,
$\eta=1.0$). In the absence of geometric dilution, we would expect
$\eta_{\rm eff} = 1.0$ and $0.5$ for radial surface emissivity
distributions ($F(r) \propto r^{\gamma}$) with power-law indices
$\gamma$ of $-2$ and $-1$ respectively.

\begin{figure}
\resizebox{\hsize}{!}{\includegraphics[angle=0,width=8cm]{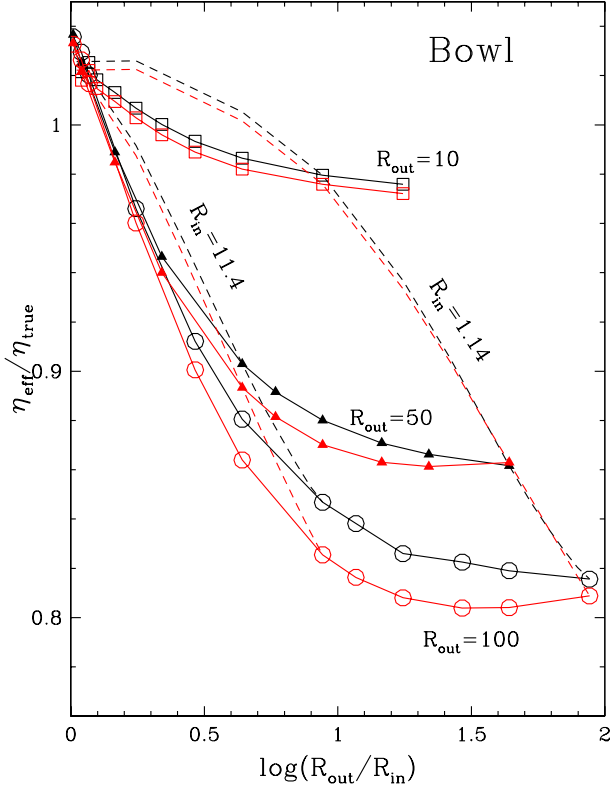}}
\caption{$\eta_{\rm eff}/\eta_{\rm true}$ as a function of
$R_{\rm out}/R_{\rm in}$ for a bowl-shaped geometry, and fixed $R_{\rm out}$ of
100 (open circles), 50 (filled triangles) and 10 (open squares)
light-days. Also shown (dashed lines), are $\eta_{\rm eff}/\eta_{\rm true}$ as
a function of $R_{\rm out}/R_{\rm in}$ for fixed $R_{\rm in}$ of 1.14, 11.4
light-days. The colours refer to the slope of the power-law emissivity
distribution (black: power-law slope $-1$, $\eta(r)=-(\gamma/2)=0.5$, red: power-law
slope $-2$, $\eta(r)=-(\gamma/2)=1.0$).}
\label{eta_bowl}
\end{figure}

\begin{figure}
\resizebox{\hsize}{!}{\includegraphics[angle=0,width=8cm]{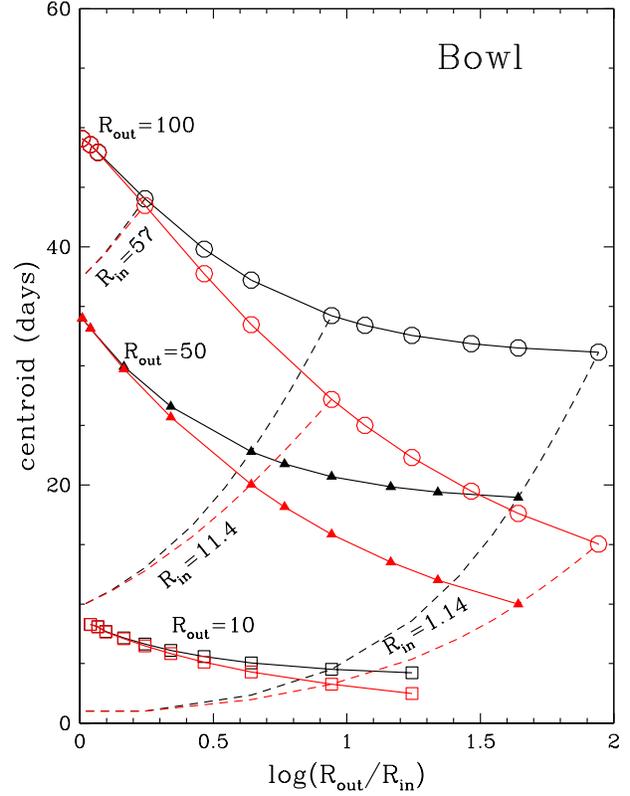}}
\caption{The centroid of the 1-d response function as a function of
$R_{\rm out}/R_{\rm in}$ for a bowl-shaped BLR geometry and fixed
$R_{\rm out}$ of 100 (open circles), 50 (filled triangles) and 10
(open squares) light-days. Also shown (dashed lines) is the centroid
of the 1-d response function as a function of $R_{\rm out}/R_{\rm in}$
for fixed $R_{\rm in}$ of 1.14, 11.4 and 57.0 light-days. The colours
refer to the slope of the radial power-law emissivity distribution
(black: power-law slope $-1$, $\eta(r)=-(\gamma/2)=0.5$, red: power-law slope $-2$,
$\eta(r)=-(\gamma/2)=1.0$).}
\label{lag_bowl}
\end{figure}

For a spherical BLR geometry the ratio $\eta_{\rm eff}/\eta_{\rm
  true}$ remains $\approx$constant for fixed $R_{\rm out}$, for
$R_{\rm out}/R_{\rm in} > 10$. For geometrically-thin shells,
ie. $R_{\rm out}/R_{\rm in} <$ a few, $\eta_{\rm eff}/\eta_{\rm true}$
declines as $R_{\rm in}$ increases. Figure~\ref{lag_sphere}, which
illustrates the variation in the luminosity-weighted radius (or
equivalently the centroid of the 1-d response function, the
responsivity-weighted radius) as a function of $R_{\rm out}/R_{\rm
  in}$, suggests that the origin of this relation is tied to the near
constancy of the responsivity-weighted radius (or response function
centroid) for large $R_{\rm out}/R_{\rm in}$, and the rapid increase
in the luminosity-weighted radius for small $R_{\rm out}/R_{\rm
  in}$. For our spherical BLR, the emission-line luminosity varies as
$dL(r) \propto r^{2+\gamma} dr$, where $\gamma$ is the power-law index
of the radial surface emissivity distribution. This yields
responsivity-weighted radii of 66 light-days and 49.5 light-days for a
BLR spanning $R_{\rm in}=1.14$ and $R_{\rm out}=100$ light-days, for
power-law indices $\gamma= -1$ and $\gamma=-2$ respectively.

Similar to Gilbert and Peterson (2003) we find that for thin-shell
geometries with outer radii $R_{\rm out} < 10$ light-days, $\eta_{\rm
eff}$ is close to 90\% of the expected value. It is also clear that
geometric dilution is larger for flatter emissivity distributions (for
a geometrically-thick BLR), and converges for geometrically-thin
regions (the separation between the black and red lines declines as
$R_{\rm out}/R_{\rm in}\rightarrow 1$). Figure~\ref{eta_sphere} also
illustrates that if the location of the BLR inner and outer boundaries
are allowed to vary with continuum level, then for a fixed radial
surface emissivity distribution, the largest variation in $\eta_{\rm
eff}$ results from changes in the location of the BLR outer boundary,
for example, follow the dashed vertically running lines in
Figure~\ref{eta_sphere}.

\subsubsection{Disc-shaped BLR geometries}

\begin{figure}
\resizebox{\hsize}{!}{\includegraphics[angle=0,width=8cm]{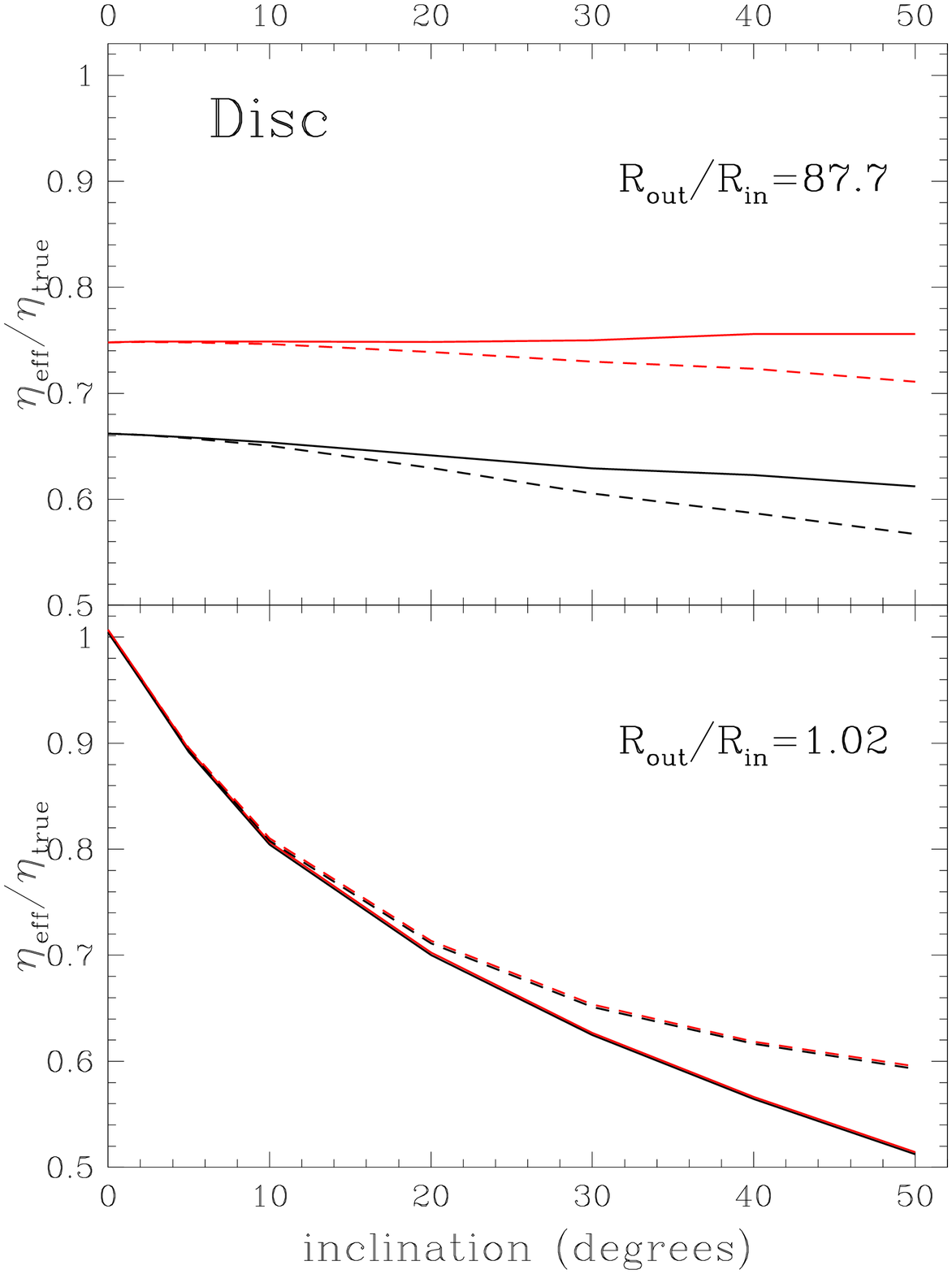}}
\caption{upper panel -- $\eta_{\rm eff}/\eta_{\rm true}$ for 
$R_{\rm out}=100$ light-days, $R_{\rm in}=1.14$ light-days, and power-law emissivity
distributions of slope $-1$, $-2$ (equivalently $\eta(r)=0.5$ (black),
1.0 (red) ). Solid lines indicate isotropic emission, dashed lines
100\% anisotropy. Lower panel -- as above for $R_{\rm
in}=98.07$~light-days (ie. $R_{\rm out}/R_{\rm in}\approx 1.0$)}
\label{anis}
\end{figure}

Figure~\ref{eta_disk} indicates the ratio $\eta_{\rm eff}/\eta_{\rm
true}$ as a function of $R_{\rm out}/R_{\rm in}$ for a face-on
disc-shaped BLR geometry. In Figure~\ref{lag_disc} we show the
corresponding dependence of the luminosity-weighted radius on $R_{\rm
out}/R_{\rm in}$ for this geometry.  As for spherical BLR geometries,
Figure~\ref{eta_disk} reveals that for spatially extended regions,
ie. $R_{\rm out}/R_{\rm in} > 10$, changes in $\eta_{\rm eff}$ due to
geometric dilution are dominated by changes in the BLR outer radius.
For example, if the BLR outer radius extends to 100 light-days, a
factor of 10 change in the inner radius (1.14--11.4 light-days) has
little effect on $\eta_{\rm eff}$ since the responsivity-weighted
radius  increases only
very slowly with $R_{\rm in}$. By comparison, for fixed $R_{\rm in}$ a
factor of 10 decrease in the outer radius can produce a significant
increase in the line responsivity ($\sim 30$\% for $R_{\rm out}/R_{\rm
in} > 10$), owing to the rapid drop in the responsivity-weighted
radius with decreasing $R_{\rm out}$ (e.g. Figure~\ref{lag_disc}).
However, unlike the spherical BLR geometry, for geometrically-thin
discs (or rings), large changes in $\eta_{\rm eff}$ for fixed
$R_{out}$ can also arise simply by increasing the inner radius. This
difference arises because for a thin shell, the
continuum--emission-line delays span the range $0 \le
\tau \le 2R/c$, while for a face-on thin ring of emitting material,
the delay is simply $R/c$. The results presented here will change for
inclined discs, though not significantly over the expected range in
line-of sight inclinations $i \le 30$ degrees (see
e.g. Figure~\ref{anis} for details). For our disc-shaped BLR, $dL(r)
\propto r^{1+\gamma}$.  The responsivity-weighted radii for a BLR
spanning $R_{\rm in}=1.14$ and $R_{\rm out}=100$ light-days is 49.5
and 22 light-days for power-law indices $\gamma=-1$, and $\gamma=-2$
respectively (e.g. Figure~\ref{lag_disc}).  Thus for a disc-shaped
BLR, assuming fixed BLR boundaries larger responsivities (e.g. $\eta_{\rm
eff}/\eta_{\rm true} > 0.9$) requires either a small outer radius
($R_{\rm out} < 30$ light-days), or a Ring-like geometry ($R_{\rm
out}/R_{\rm in} <$ a few).

\subsubsection{Bowl-shaped BLR geometries}

Figures~\ref{eta_bowl} and \ref{lag_bowl} indicate the equivalent
relationships for our fiducial bowl-shaped BLR described in \S3.1 (see
also Goad, Korista and Ruff 2012), this time when viewed face-on
($i=0$).  In this model, the BLR bridges the gap between the outer
accretion disc and the inner edge of the dusty torus via an effective
surface surface the scale height of which increases with increasing
radial distance. By placing material at larger radial distances closer
to an observer's line of sight, geometric dilution of the line
responsivity is significantly reduced, due to the reduced spread in
delays for this geometry, up to a factor 2 for $R_{\rm in}=1.14$
light-days and $R_{\rm out}=100$ light-days (note that for a bowl-like
geometry, the gas at larger radii has a smaller surface area for the
same radial distance than for a disc), though the general dependence
on $R_{\rm out}$ and $R_{\rm in}$ is, by and large, the same as for
the thin-disc geometry\footnote{When taken to the extreme, where the
  contours of the bowl follow a parabolic surface, the delay will be
  the same everywhere (an iso-delay surface), regardless of where a
  line forms.}. The responsivity-weighted radii for our bowl-shaped
geometry when viewed face-on and spanning BLR radii $R_{\rm in}=1.14 $ --
$R_{\rm out}=100$ light-days are 31.1 and 15.0 light-days for
power-law indices $\gamma=-1$, and $\gamma=-2$ respectively.

\begin{figure}
\resizebox{\hsize}{!}{\includegraphics[angle=0,width=8cm]{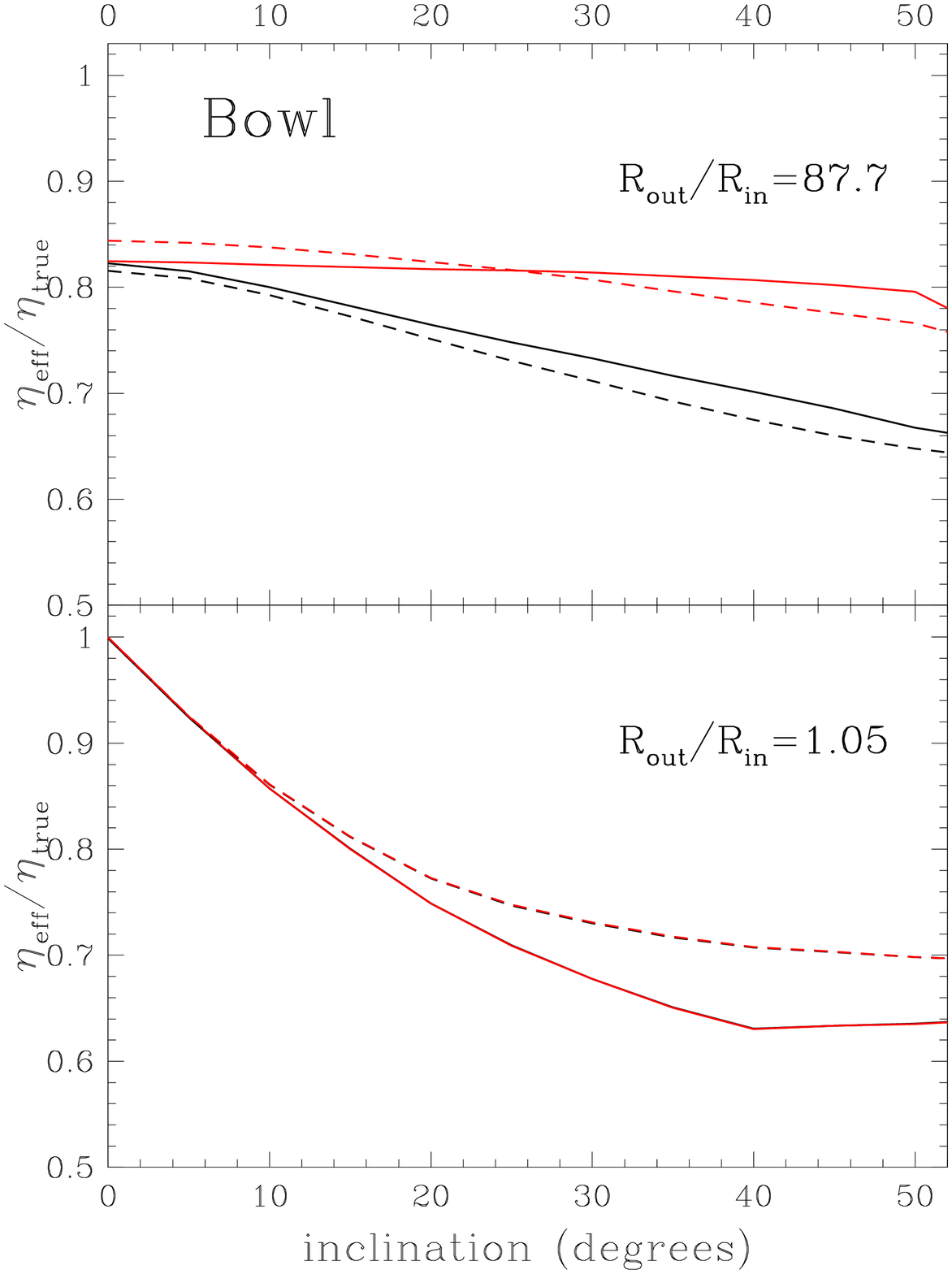}}
\caption{upper panel -- $\eta_{\rm eff}/\eta_{\rm true}$ for 
$R_{\rm out}=100$ light-days, $R_{\rm in}=1.14$ light-days, and power-law emissivity
distributions of slope $-1$, $-2$ (equivalently $\eta(r)=0.5$ (black),
1.0 (red) ). Solid lines indicate isotropic emission, dashed lines
100\% anisotropy. Lower panel -- as above for $R_{\rm
in}=98.07$~light-days (ie. $R_{\rm out}/R_{\rm in}\approx 1.0$)}
\label{anis_bowl}
\end{figure}

\subsection{Anisotropy and inclination}

The disc- and bowl-shaped BLR geometries presented in \S3.3 assume
isotropic cloud emission for a BLR which is viewed face-on
(ie. $i=0$). We have also investigated spherical, thin-disc and
bowl-shaped geometries in which the line emission from individual BLR
clouds is 100\% inward toward the ionising continuum source
(i.e. fully anisotropic emission), as well as exploring the variation
in geometric dilution with line-of-sight inclination. For the
anisotropy we here adopt a form which approximates the phases of the
moon (e.g. Goad 1995; O'Brien, Goad and Gondhalekar 1994). The effect
of anisotropy on the responsivity-weighted radius for both spherical
and disc-like BLR geometries has been explored elsewhere
(e.g. O'Brien, Goad and Gondhalekar 1994; Goad 1995, PhD thesis).

For geometrically-thick spherical BLR geometries, anisotropy enhances
the far side emission relative to the near side emission. The net
result is that the response of gas lying closest to the line of sight
and which responds on the shortest timescales decreases significantly
as the anisotropy increases and consequently the mean response
timescale as given by the responsivity-weighted radius (or
equivalently the luminosity-weighted radius) increases. In terms of
the measured responsivity, the larger responsivity-weighted radius
will result in a smaller amplitude emission-line response.

A similar outcome applies to disc-like BLR geometries, though for our
adopted form of anisotropy the responsivity-weighted size of the
emitting region has a strong inclination dependence, increasing as the
inclination increases (note that increased emission-line anisotropy
has no effect on the mean response timescale at zero inclination for a
disc).  Figure~\ref{anis} (upper panel) illustrates the effect of line
anisotropy and inclination on the ratio $\eta_{\rm eff}/\eta_{\rm
  true}$ for a thin-disc geometry, $R_{\rm out}=100$~light-days,
$R_{\rm in}=1.14$ light-days and line-of-sight inclination 0--50
degrees.  Solid lines indicate isotropic line emission, dashed lines
100\% anisotropy, while colours indicate $\eta=0.5$ (black),
$\eta=1.0$ (red).  For geometrically-thick ($R_{\rm out}/R_{\rm in} >
10$) disc-shaped BLR geometries anisotropy and inclination produce
only a modest change in the effective responsivity ($<10$\% even at
the largest inclination). For a disc-like geometry with purely
isotropic emission, inclination increases the spread in the
emission-line delays, but the responsivity-weighted size of the BLR
remains unchanged. When anisotropy is included, the
responsivity-weighted size of a geometrically-thick disc-like BLR
increases only very slowly with increasing inclination, because
significant front-back asymmetry occurs only for relatively large
inclinations.

By contrast, for geometrically-thin disc-like BLR geometries ($R_{\rm
  out}/R_{\rm in} \sim 1$) the line responsivity $\eta_{\rm eff}$
decreases dramatically with increasing inclination, a purely
geometrical effect, while anisotropy tends to increase the
responsivity relative to the isotropic case at large
inclinations. Though counter intuitive, the latter arises because
while the responsivity-weighted size of the BLR increases with
increased inclination (in the anisotropic case), the spread in delays
is smaller for a thin ring with anisotropic emission relative to the
purely isotropic case. For example, if we adopt 30~degrees as a
typical inclination, then for a thin ring with $R_{\rm out}/R_{\rm in}
\approx 1$, the measured responsivity relative to that found for a
face-on thin ring of the same dimensions decreases by $\sim 38$\% for
isotropic emission, and only $\sim 34$\% for fully anisotropic
emission (see e.g. Figure~16).

Our fiducial bowl-like BLR geometry exhibits similar behaviour with
anisotropy and inclination to that of a disc-like BLR, but with some
subtle geometry related differences (Figure~\ref{anis_bowl}).  The
bowl is not front-back symmetric, thus even in the isotropic emission
case, the responsivity-weighted radius will increase with increasing
inclination, though only weakly. However, at inclinations greater than
45 degrees self-occultation occurs and the delay increases
significantly as substantial material lying close to the line of sight
is obscured from view (see e.g. Figure~1 of Goad, Korista and Ruff
2012). When anisotropy is included, initially the mean delay is
reduced as emission from gas lying on the steep sides of the bowl is
reduced, removing contributions of gas at large delays.  As the bowl
is inclined further, the responsivity-weighted radius increases. When
considering only the extreme outer rim of the bowl, the observed
behaviour is similar to that for a thin ring, $\eta_{\rm eff}$
decreases sharply with increased inclination.  However, for the bowl
the responsivity stabilises for inclinations above $\approx 20$
degrees.

\section{Discussion : a Breathing BLR}\label{discuss}

The 13~yr ground-based spectroscopic monitoring campaign of the nearby
Seyfert 1 galaxy NGC~5548 indicates that the broad H$\beta$ line EW is
time-variable, changing by more than a factor of 2 on timescales less
than a typical observing season ($\approx$~200 days) (lower panel of
Figure~1). Indeed the measured broad H$\beta$ line EW, which is a
measure of the continuum re-processing efficiency of the line-emitting
gas for this line, is found to be inversely correlated with continuum
level, such that the largest EWs are found during low continuum
states. The line EW is related to the effective emission-line
responsivity $\eta_{\rm eff}$ via equation~2, which indicates that a
time-variable line EW is to be expected if $\eta_{\rm eff} \neq 1$. If
even after accounting for the effects of geometric dilution $\eta_{\rm
  eff} \approx 1$, then following equation~2 the re-processing
efficiency for a particular line will remain approximately constant
(ie. EW(line)$\approx$ constant, ie. no intrinsic Baldwin effect).
Since EW(H$\beta$) is found to be inversely correlated with continuum
level in NGC~5548 (an intrinsic Baldwin effect for H$\beta$), the
effective responsivity must be less than $1$ for this line.

We note here that the inverse correlation found between ionising
continuum flux and the emission-line re-processing efficiency is a key
prediction (and a major success) of photoionisation model calculations
(e.g. Korista and Goad 2004, their Figures 3 and 4). Specifically,
Korista and Goad (2004) found that for the hydrogen and helium
recombination lines the local (in-situ) emission-line responsivity
$\eta$ increases toward lower hydrogen ionising continuum fluxes
independent of any assumed geometry. Thus in the absence of those
effects which act to reduce the emission-line responsivity
(e.g. geometric dilution), photoionisation models predict that locally
the emission-line responsivity $\eta$ will increase as the ionising
continuum flux decreases (ie. $\eta = \eta(L_{\rm cont}(t))$,
e.g. Figure~2, cf. solid and dashed black lines).

A key goal of the present work was to determine how the properties of
the driving continuum light-curve (ie. its amplitude and
characteristic timescale), observational bias (e.g. light-curve
duration and sampling pattern) and choice of BLR geometry, contribute
to the measured emission-line responsivity $\eta_{\rm eff}$.  To
isolate these effects from those caused by changes in the local gas
physics arising from ionising continuum flux variations, we have
assumed a radial surface emissivity distribution which is a power-law
in radius and a locally-linear response approximation for a BLR with
fixed inner and outer boundaries.  For a power-law radial surface
emissivity distribution, $F(r)\propto r^{\gamma}$,
$\eta(r)=-(\gamma/2)=constant\, , \forall r$. While locally the radial
surface emissivity increases with continuum level, it does so by the
same amount everywhere. Thus the radial dependence, if not the
absolute value of the radial surface emissivity distribution, is the
same and thus the instantaneous responsivity-weighted radius is
constant in time. Since both the radial line responsivity and
responsivity-weighted radius are constant $\forall r,t$, such a
broad-line region cannot breathe (the characteristic emission-line
response amplitude and delay are constant in time). Thus for these
models we do not expect to find a correlation between BLR size and
continuum state nor an inverse correlation between the emission-line
responsivity $\eta_{\rm eff}$ and continuum state.

Photoionisation model calculations by Goad, O'Brien and Gondhalekar
(1993), and Korista and Goad (2004), indicate that power-law radial
surface emissivity distributions are generally a poor approximation to
the predicted behaviour of the majority of the strong broad UV and
optical emission-lines.  In Figure~2 we illustrate the radial surface
emissivity distribution $F(r)$ (upper panel) and radial line
responsivity $\eta(r)$ (lower panel) for broad H$\beta$ as determined
for an LOC model of the BLR in NGC~5548 (Korista and Goad 2004).  

For broad H$\beta$ the radial surface emissivity distribution is best
represented by a broken power-law with a power-law index $\gamma
\approx -0.7$ for radii less than $\approx 25$ light-days and breaks
towards values in $\gamma$ steeper than $\approx -2$ for radii greater
than $\approx 160$ light-days (as indicated by the red line in the
upper panel of Figure~2). The corresponding values in the radial
responsivity distribution spans $\eta(r) \approx 0.35$ in the inner
BLR to values of $\eta(r) \approx 1$ or more beyond the confines of
our fiducial BLR outer boundary.

Because in general, emission-line emissivities are not strictly
power-laws, $\eta(r) \neq constant$, such a BLR will breathe even if
the boundaries of the BLR remain fixed at their starting values
(Korista and Goad 2004). Consequently the responsivity-weighted radius
(or equivalently the luminosity-weighted radius as measured from the
centroid of the CCF) will vary with continuum state due to changes in
the relative radial surface emissivity distribution with continuum
flux within the confines of those fixed boundaries. Figure~2
illustrates the H$\beta$ radial surface emissivity distributions and
their respective radial responsivity distributions corresponding to
small continuum variations about low- and high-continuum states
(dashed and solid black lines respectively).  These states were chosen
to match two historical extrema in the UV continuum flux of NGC~5548
and correspond to a peak-peak variation of $\approx 8.2$ (or
equivalently $0.5\times \log 8.2=0.457$ in $\log r$).
Compare their behaviour to that of a simple power-law radial
emissivity distribution (as indicated by the green lines in Figure~2)
for which the luminosity-weighted radius and emission-line
responsivity are invariant.

For most lines, including H$\beta$, the local radial responsivity
distribution $\eta(r)$ generally decreases with increased ionising
continuum flux, which when integrated over the whole BLR results in a
reduction in the emission-line responsivity and an
increase in the delay (e.g. Korista and Goad 2004, their Figure~3;
Goad, Korista and Ruff 2012, their Figure~9). This finding is
independent of those effects discussed in \S3 (for example, $T_{\rm
  char}$, $T_{\rm dur}$ and BLR geometry) which when folded in, act to
further reduce the measured emission-line responsivity.  Korista and
Goad (2004) found low- and high-continuum state responsivities for
H$\beta$ spanning the range 0.54--0.77 (columns 3 and 4 of their
Table~1) in the absence of those effects which act to reduce the
emission-line responsivity. The measured range in $\eta_{\rm eff}$
when referenced to the amplitude of the UV continuum variations is
significantly larger, $0.33 < \eta_{\rm eff} < 0.84$, and is inversely
correlated with continuum level (Goad, Korista and Knigge 2004; Bentz
et al. 2009).

Figures~\ref{eta_disk} and \ref{eta_bowl} show that for steep radial
surface emissivity distributions, it is difficult for geometric
dilution alone to reduce $\eta_{\rm eff}$ to values as low as $\approx
0.3$. With our adopted value of $T_{\rm char} = 40$~days the
emission-line response is already significantly geometrically diluted
for the geometries presented here. By implication, the radial surface
emissivity distribution for H$\beta$ is unlikely to be steep as
$\gamma=-2$.

The measured upper bound of $\eta_{\rm eff}=0.84$ is larger than the
upper bound quoted in Korista and Goad (2004). This is particularly
intriguing, since we have already shown that $T_{\rm char}$, $T_{\rm
  dur}$ and BLR geometry, generally act to dilute the emission-line
responsivity.  Adopting a smaller outer boundary for our model BLR
cannot help here because the in-situ responsivity is small at small
BLR radii for broad H$\beta$.  One possible explanation is that the
amplitude of the driving ionising continuum is significantly larger
than that used here, and subsequently the measured range in
emission-line responsivity when integrated over the whole BLR will be
larger between the high- and low- continuum states than considered in
Korista and Goad (2004).  For a BLR of fixed radial extent we can
enhance the radial responsivity at smaller BLR radii by adopting a
smaller continuum normalisation. This will consequently reduce the
responsivity-weighted radius. When reverberation effects are then
taken into consideration, this will produce a more coherent
emission-line response.  A smaller continuum normalisation may also
result in a smaller BLR outer boundary.

Because of geometric dilution, spatially extended BLRs will exhibit a
large effective responsivity in H$\beta$ only if the intrinsic (local
gas) responsivities are significantly larger than those shown
here. Such large intrinsic responsivities require low incident
ionising photon fluxes, which as shown in Figure~2 lie beyond the
outer boundary of our model geometry.  Alternatively, larger
$\eta_{\rm eff}$ for H$\beta$ may also be obtained if the radial
surface emissivity curve is moderately steeper than that shown
here. This can be achieved for example by including gas at densities
$> 10^{12}$~cm$^{-3}$, or if locally the line emitting gas experiences
significant extra-thermal line broadening (e.g. Bottorff et al. 2000).

Finally, we haven't yet touched upon the possibility that the location
of the BLR inner and outer boundaries may also vary in response to
continuum variations.  Figures~10--15 suggest that relaxing the
assumption of fixed BLR boundaries will allow large changes in the
emission-line responsivity and delay to occur.  We defer investigation
of the effects of a radially-dependent and continuum-level dependent
emission-line responsivity (breathing) and the possible time-variable
location of the BLR inner and outer boundaries in response to
continuum variations to paper~{\sc ii}.


\section{Conclusions}

The measured amplitude and delay of the emission-line response to
continuum variations is determined by the local gas physics, the BLR
geometry and the characteristics of the driving ionising continuum
($T_{\rm char}$, $T_{\rm dur}$ and $\Delta t$). The role of the local
gas physics in determining the emission-line responsivity has been
explored in Korista and Goad (2004). Here, we focus on the remaining
aforementioned effects which act to modify the local emission-line
responsivity.

We have demonstrated two independent methods for the determination of
the effective emission-line responsivity $\eta_{\rm eff}$ each with
similar accuracy.  The standard approach using the logarithmic slope
of the relation between the continuum--emission-line fluxes $d \log
f_{\rm line}/d \log f_{\rm cont}$, is more difficult to determine,
requiring a correction for the continuum--emission-line delay and
normally involves interpolating on one of the light-curves. The second
method is far simpler to implement and requires only the determination
of the ratio of the fractional variance of the line to the fractional
variance of the continuum $F_{\rm var}$(line)/$F_{\rm
  var}$(cont). Indeed for the latter, continuum and emission-line data
need not be contemporaneous, though the dependence of responsivity on
continuum level is lost (the latter may in part be recovered by
investigating the variability of the continuum and emission-line
light-curves as a function of continuum state).  Both methods require
a suitable correction for contaminating non-variable components in the
line (from contaminating narrow mission-lines) and continuum (from the
host galaxy).

Importantly, our simulations indicate that light-curve duration plays
an important role in the determination of the effective emission-line
responsivity $\eta_{\rm eff}$, and a similar effect is found for the
continuum emission-line delay.  For short observing campaigns the
measured value of $\eta_{\rm eff}$ is biased towards smaller values
with a larger intrinsic scatter.  Reduced sampling rates do not affect
the measured value of $\eta_{\rm eff}$, though the error on a single
estimate increases as the sampling rate decreases.  This suggests that
measuring the ``true'' responsivity of a given emission-line requires
light-curves of sufficient duration, typically longer than the
autocorrelation function of the continuum light-curve. While
noteworthy, the observational bias toward smaller responsivity
introduced by a shorter duration campaigns can not explain the
observed inverse correlation between $\eta_{\rm eff}$ and continuum
level. Simulations such as these may be used to inform the design of
future intensive reverberation mapping campaigns.

Contrary to previous work, our simulations indicate that geometric
dilution may play a significant role in reducing the effective
emission-line responsivity.  For a BLR of fixed spatial extent, and a
characteristic continuum variability timescale $T_{\rm char}$ which is
less than the maximum delay for a particular line $\tau_{\rm
  max}$(line), for a given BLR geometry and observer line-of-sight
orientation, the effective emission-line responsivity $\eta_{\rm
  eff}$, emission-line delay and the maximum of the
continuum--emission-line cross-correlation coefficient are strongly
correlated (Figure 9, panels (ii) and (v)).  For example, our
simulations suggest that for our fiducial BLR model, $\eta_{\rm eff}$
decreases significantly for $T_{\rm char} \le 100$~days, the maximum
delay at the outer radius for our fiducial BLR geometry when observed
at a line of sight inclination of 30~degrees. This suggests that for a
given BLR geometry, the characteristic continuum variability
timescale, $T_{\rm char}$, is a key quantity in the determination of
the measured emission-line responsivity and delay. Importantly, if
$T_{\rm char} \ge \tau_{\rm max}$(line) geometric dilution is
minimised and the measured responsivity and delay may be considered as
representative of the underlying gas responsivity and mean response
timescale.  Conversely, if $T_{\rm char} < \tau_{\rm max}$(line),
geometric dilution can act to significantly reduce the measured
emission-line responsivity and delay.

Changes in the short timescale variability amplitude of the continuum
light-curve appear to be less important in the determination of
$\eta_{\rm eff}$ and delay, over the range in variability amplitude
expected for the driving continuum light-curve.  Significantly, if for
a fixed continuum luminosity, the characteristic continuum variability
timescale $T_{\rm char}$ and BLR 'size' varies among the AGN
population, then geometric dilution in individual sources will
introduce scatter into the well-known BLR radius--luminosity relation.

Following on from the work of Gilbert and Peterson (2003), we have
explored the effect of geometric dilution on emission-line
responsivity for spherical, disc and bowl-shaped BLR geometries,
spanning a range of BLR sizes (both geometrically-thick and
geometrically-thin) using a fixed characteristic timescale and
amplitude for the driving continuum light-curve.  In the majority of
cases the measured responsivities are found to be largest for smaller
(more compact) BLRs (Figures 10, 12, and 14). For disc- and
bowl-shaped BLR geometries, the largest responsivities are found for
geometrically-thin BLRs, neglecting the effects of inclination.  For
geometrically-thick BLRs, the measured responsivity $\eta_{\rm eff}$
is strongly correlated with the characteristic size of the
line-emitting region (Figures 11, 13, and 15).  Significantly, Figures
10, 12, and 14 suggest that large changes in an emission-line's
responsivity may be realised, if we relax the assumption of fixed BLR
boundaries. We explore this further in paper~{\sc ii}.

\section{Acknowledgements}

\thanks{We would like to thank the referee for some helpful
  suggestions which have helped to improve the clarity of the work
  presented here. We would also like to thank Anna Pancoast for a very
  careful reading of the original manuscript which helped to improve
  the clarity of the work presented here. Mike Goad would also like to
  thank Kirk and Angela Korista and Western Michigan University for
  their generous hospitality during the early stages of this work.}

\section{References}


\noindent Baldwin, J. Ferland, G., Korista, K., and Verner, D., 1995, ApJ 455, 119.



\noindent Bentz, M.C., Walsh, J.L., Barth, A.J., Baliber, N., Bennert, V.N. et al., 2009, ApJ 705, 199.


\noindent Bentz, M.C., Peterson, B.M., Pogge, R.W. et al., 2006, ApJ 644, 133.

\noindent Bottorff, M., Ferland, G., Baldwin, J. and Korista, K. 2000, ApJ 542, 644.

\noindent Cackett, E.M. and Horne, K., 2006, MNRAS 365, 1180.

\noindent Clavel, J. Reichert, G.A. Alloin, D. et al., 1991, ApJ 366, 64.


\noindent Collier, S. and Peterson, B.M., 2001, ApJ 555, 775.

\noindent Collier, S.J., Peterson, B.M. and Horne, K., 2001, BAAS, vol 333, p896.

\noindent Czerny, B., Doroshenko, V. T., Nikołajuk, M., et al., 2003, MNRAS, 342, 1222 

\noindent Denney, K.D., Peterson, B.M., Dietrich, M., Vestergaard, M. and Bentz, M.C., 2009, ApJ 692, 246.

\noindent Ferland, G.J., Peterson, B.M., Horne, K. Welsh, W.F., Nahar, S.N., 1992, ApJ 387, 95.

\noindent Gaskell, C.M. and Peterson, B.M., 1987, ApJS 65, 1.

\noindent Gilbert, K.M. and Peterson, B.M., 2003, ApJ 587, 123.

\noindent Goad, M. R.; Korista, K. T. and Ruff, A.J., 2012, MNRAS 426, 3086.

\noindent Goad, M. R.; Korista, K. T.; Knigge, C., 2004, MNRAS 352, 277.

\noindent Goad, M.R. and Wanders, I., 1996, ApJ 469, 113.

\noindent Goad, M.R., 1995, PhD thesis, University College London.

\noindent Goad, M.R., O'Brien, P.T. and Gondhalekar, P.M., 1993, MNRAS 263, 149.


\noindent Han, X.-H., Wang, J., Wei, J.-Y., Yang, D.-W. and Hou, J.-L., 2011,  Science China, Physics Mechanics and Astronomy, 54, 346.

\noindent Horne, K., Welsh, W.F. and Peterson, B.M. 1991, ApJ 367, L5.

\noindent Horne, K., Peterson, B.M., Collier, S.J. and Netzer, H., 2002, astro-ph 1182.

\noindent Horne, K., Korista, K.T. and Goad, M.R., 2003, MNRAS 339, 367.

\noindent Kong, M.-Z., Wu, X.-B, Wang, R. Liu, F.K. and Han, J.L., 2006, A\&A 456, 473.


\noindent Kelly, B.C., and Bechtold, J., 2007 ApJS 168, 1.

\noindent Kelly, B.C., Bechtold, J., Siemiginowska, A., 2009, ApJ 698, 895.

\noindent Koratkar, A.P. and Gaskell, C.M. ApJS  74, 719.

\noindent Korista, K.T., Alloin, D., Barr, P. et al., 1995, ApJS 97, 258.

\noindent Korista, K.T. and Goad, M.R., 2000, ApJ 536, 284.

\noindent Korista, K.T. and Goad, M.R., 2001, ApJ 553, 695.

\noindent Korista, K.T. and Goad, M.R., 2004, ApJ 606, 749.

\noindent Kozlowski, S. Kochanek, C.S., Udalski, A. et al., 2010, ApJ 708, 927.

\noindent Krause, M., Schartmann, M. and Burkert, A., 2012, MNRAS 425, 3172.

\noindent Krolik, J.H., Horne, K., Kallman, T.R. et~al., 1991, ApJ 371, 541.

\noindent MacLeod, C.L., Ivezi\'{c}, Z., Kochanek,C.S. et al., 2010, ApJ 721, 1014.


\noindent Maoz, D., Netzer, H., Peterson, B.M., Bechtold, J., Bertram, R. et al., 1993, ApJ 40, 576.

\noindent Maoz, D. 1992 In ``Physics of Active Galactic Nuclei.'' Editors, W.J. Duschl, S.J. Wagner, P.214, 1992


\noindent Netzer, H. and Maoz, D., 1990, ApJ, 365, 5.

\noindent Netzer, H. and Laor, A., 1993, ApJ 404, 51.


\noindent O'Brien, P.T. Goad, M.R. and Gondhalekar,  P.M., 1994, MNRAS 268, 845.

\noindent O'Brien, P.T., Goad, M.R. and Gondhalekar, P.M., 1995, MNRAS, 275, 1125.


 
\noindent Penston, M. V. 1991, in ``Variability of Active Galactic Nuclei'' eds, Miller H.R., Wiita, P.J. Cambridge University Press, p43.

\noindent Peterson, B. M., Berlind, P., Bertram, R., Bischoff, K., Bochkarev, N. G et al., 2002, ApJ 581, 197.

\noindent Peterson, B.M., Alloin, D., Axon, D., Balonek, T.J., Betram, R. et al., 1992, ApJ 392, 470.

\noindent Peterson, B. M., Berlind, P., Bertram, R., Bochkarev, N. G., Bond, D., et al., 1994, ApJ 425, 622.

\noindent Peterson, B.M., Denney, K.D., De Rosa, G., Grier, C.J., Pogge, R.W. et al., 2013, ApJ 779, 109.

\noindent Pogge, R.W. and Peterson, B.M., 1992, AJ, 103, 1084.

\noindent P\'{e}rez, E., Robinson, A., and de La Fuente, L., 1992a, MNRAS 255, 502.

\noindent P\'{e}rez, E., Robinson, A., and de La Fuente, L., 1992b, MNRAS 256, 103.

\noindent Press, W.H., Teukolsky, S.A., Saul, A., Vetterling, W.T. and Flannery, P., 1992. Cambridge: University Press, c1992, 2nd ed.

\noindent Robinson, A. and P\'{e}rez, E., 1990. MNRAS, 244, 138.


\noindent Runnoe, J.C., Brotherton, M.S., Shang, Z., Wills, B.J. and  DiPompeo, M.A., 2013, MNRAS 429, 135.


\noindent Shen, Y. Greene, J.E. Strauss, M.A., Richards, G.T. and Schneider, D.P., 2008, ApJ 680, 196.

\noindent Shen, Y. and Kelly, B.C. 2010, ApJ 713, 41.


\noindent Suganuma, M., Yoshii, Y., Kobayashi, Y. et al., 2006, ApJ 639, 46.



\noindent Uttley, P., McHardy, I.M., and Vaughan, S., 2005, MNRAS 359, 345.

\noindent Vestergaard, M. and Peterson, B.M., 2005, ApJ 625, 688.



\noindent Welsh, W.F., 1999, PASP 111, 1347.

\noindent Welsh, W.F. and Horne, K., 1991, ApJ 379, 586.

\noindent Zu, Y., Kochanek, C.S., Peterson, B.M., 2011, ApJ 735, 80.

\appendix
\label{lastpage}

\end{document}